%% file: main.tex
\ifdefined\XeTeXrevision\else\pdfoutput=1\fi
\documentclass{article}

\usepackage[T1]{fontenc}

\usepackage{microtype}
\usepackage{graphicx}
\usepackage{booktabs}
\usepackage{hyperref}

\usepackage[preprint]{icml2026}

\usepackage{amsmath}
\usepackage{amssymb}
\usepackage{multirow}
\usepackage{enumitem}
\usepackage[capitalize,noabbrev]{cleveref}

\hypersetup{pdfsubject={}}

\graphicspath{{figures/}}
\setlist{itemsep=1pt,topsep=2pt,parsep=0pt,leftmargin=*}

\newcommand{\geneblind}{GeneBlind}
\newcommand{\evo}{Evo~2}
\newcommand{\esmc}{ESMC}
\newcommand{\qfour}{Qwen3.5-4B}
\newcommand{\qtwentyseven}{Qwen3.6-27B}
\newcommand{\mistral}{Mistral-7B}
\newcommand{\olmo}{OLMo-2-7B}
\newcommand{\qbig}{Qwen3.8-27B}
\newcommand{\dsflash}{DeepSeek-V4-Flash}

\newcommand{\ci}[2]{[#1,\,#2]}
\newcommand{\arm}[1]{\textsc{#1}}
\newcommand{\fpath}{\textsc{Pathway}}
\newcommand{\fdis}{\textsc{Disease}}
\newcommand{\fblock}{\textsc{Blocking}}
\newcommand{\fshare}{\textsc{Shared-process}}
\newcommand{\devtune}{dev-tune}
\newcommand{\devconf}{dev-confirm}
\newcommand{\Devtune}{Dev-tune}
\newcommand{\Devconf}{Dev-confirm}

\icmltitlerunning{Inter-Model Semantic Hub: Training-Free Latent Transfer}

\begin{document}

\twocolumn[
  \icmltitle{Inter-Model Semantic Hub: Training-Free Latent Transfer from a DNA Model into Frozen Language Models}

  \begin{icmlauthorlist}
  \icmlauthor{Arash Fatehi}{ibmi,cmmc}
\end{icmlauthorlist}
\icmlaffiliation{ibmi}{Institute for Biomedical Informatics, Faculty of Medicine and University Hospital Cologne, University of Cologne, Germany}
\icmlaffiliation{cmmc}{Center for Molecular Medicine Cologne (CMMC), University of Cologne, Faculty of Medicine and University Hospital Cologne, Cologne, Germany}

  \icmlkeywords{multimodal language models, DNA foundation models, model stitching,
    activation patching, closed-form maps, interpretability}

  \vskip 0.3in
]

\csname icml@noticeprintedtrue\endcsname
{\let\thefootnote\relax\footnotetext{\hspace*{-\footnotesep}\textsuperscript{1}Institute for Biomedical Informatics, Faculty of
Medicine and University Hospital Cologne, University of Cologne, Germany.
\textsuperscript{2}Center for Molecular Medicine Cologne (CMMC), University of Cologne, Faculty of Medicine
and University Hospital Cologne, Cologne, Germany. Correspondence to: Arash Fatehi
\textless{}fatehi.arash@proton.me\textgreater{}.

\ \\
\csname Notice@String\endcsname}}

\input{sections/00-abstract}
\input{sections/01-introduction}
\input{sections/02-related-work}
\input{sections/03-method}
\input{sections/04-results}
\input{sections/06-discussion}

\section*{Impact Statement}
\input{sections/07-impact}

\newpage
\appendix
\onecolumn
\input{sections/A-appendix}

\end{document}

%% file: sections/00-abstract.tex
\begin{abstract}
The semantic hub hypothesis holds that a language model shares one internal representation across
languages and modalities; the Platonic representation hypothesis, that the representations of different
models converge. We extend them: latent-level information can be transferred into a language model's hub
from another model of another modality, with neither model trained. We test this between a DNA model and
language models that have only read text. A single linear
map, fitted in closed form, carries the DNA model's latent representation of a gene into the language
model's own latent representation of that gene's name and description; written into the language model's
hidden states, it lets the model answer questions about genes it sees only as DNA. In a single
pre-registered test on four language models, accuracy on five-option questions about held-out genes rises
from chance (20\%) to 37--46\%, while shuffled DNA and every other control stay at chance, and the transfer
is comparable to a trained connector. A second map of the same kind, fitted on text alone, carries the
transferred information on from one language model to another at a cost below 1\%. What crosses over is
the gene's identity, after which the language model answers from what it already knows; the models cannot
read the same DNA as text. Exploratory checks show that the transfer follows the DNA string rather than the
protein, reaches yeast, plant and bacterial genes with no human relative, and across modalities needs more
than the rotation that suffices between two language models.
\end{abstract}

%% file: sections/01-introduction.tex
\section{Introduction}
\label{sec:intro}

Large language models know a lot about genes: their names, functions and pathways, and the diseases
they are involved in. DNA foundation models such as \evo{} \citep{Brixi2025-Evo2} encode sequence.
Combining the two could give models that reason in language about a sequence without being told its
name. Current systems such as ChatNT \citep{Richard2024-ChatNT} and BioReason
\citep{Fallahpour2025-BioReason} train a connector from the DNA model into the LLM, and often
fine-tune the LLM itself. This training is costly and must be repeated for every pair of models. It
also leaves open what the LLM actually receives: information about the sequence, or a pointer to
knowledge it already has.

We ask what is possible with no connector training at all, as a test of two hypotheses about shared
representations. The semantic hub hypothesis \citep{Wu2024-SemanticHubHypothesis} holds that an LLM maps
semantically equivalent inputs from different languages and modalities to nearby states in its
intermediate layers \citep[see also][]{Wendler2024-LlamasWorkInEnglish}, and that it uses this shared space:
interventions in it change the model's outputs. The Platonic representation hypothesis
\citep{Huh2024-PlatonicRepresentation} holds that the representations of different models, including models
of different modalities, converge toward a shared model of the world. Together they suggest a stronger
claim, which this paper states and tests: the hub is open to \emph{other} models. Information held by
another model, of another modality, can be written into an LLM's hub by mapping it in closed form onto the
LLM's own representation of the same entity, and the LLM then uses what it already knows about that
entity, with neither model trained. We call this \emph{latent transfer}. DNA is a demanding test: the
LLMs take no DNA input of their own, and the DNA model never saw text. We fit one ridge map from frozen \evo{} features to the host's mean
representation of the gene's symbol, protein name and description at one layer. At inference, we
replace the hidden state at the DNA positions with the map's output (\cref{sec:method}). We call this
procedure the \emph{socket} and the fitted map the \emph{plug}.

The socket works. We froze the analysis plan and the layers before scoring, and evaluated the frozen
plugs once on a held-out test set whose items are split from training by homology and whose wrong
options are chosen without looking at the answer (answer-blind). Over shuffled DNA, the socket adds
$+22.2$\% on a 4B host, $+26.1$ on a 27B host, $+16.7$ on \mistral{} and $+18.4$ on \olmo{}; all
four survive Holm correction, and confirmatory development results agree. Every null control
(permuted genes, swapped genes, random maps, a zero write) stays at the sham level. The socket is
comparable to a trained MLP adaptor, without its failed seeds. Plugs move between hosts through a
linear map fitted on gene texts only, and a protein encoder, \esmc{} \citep{Candido2026-ESMC}, plugs into
the same sockets.

The socket carries identity: swapping genes removes the effect and retrieving the nearest training gene
gives most of the gain, yet the plug beats this retrieval on every host, works for non-human genes without
human relatives, and the hosts cannot read the sequence as letters. It lets the host use the knowledge it
has; it cannot add knowledge the host lacks.

\paragraph{Contributions.} Replacing a hidden state at a placeholder and continuing the forward pass
is prior art, and so are closed-form linear maps between LLMs (\cref{sec:related}). Our contributions are:
\begin{enumerate}
  \item Latent transfer, an extension of the semantic hub hypothesis to other models: a frozen LLM uses
  information from a model of another modality once it is written into the LLM's own latent
  representation of the entity, mapped in closed form, with neither model trained (\cref{sec:method,sec:results}).
  \item A closed-form plug that maps a frozen encoder of a \emph{foreign modality} into the host's
  \emph{own text space} at an early layer, with no gradient through the host. We test it under a
  frozen plan on four hosts from three families, in a single run on a held-out test set that
  follows confirmation on development data (\cref{sec:res-test,sec:res-socket}).
  \item A rule for choosing the socket layer by slot readability. It uses no task items and predicted
  the layer of a held-out host. We also document the failure of an earlier rule (\cref{sec:res-layer}).
  \item Modularity across hosts and modalities: through text-only maps, one DNA plug feeds four hosts
  from three families, and a protein plug uses the same sockets (\cref{sec:res-modularity,sec:res-protein});
  and a comparison with a closed-form rotation, which matches ridge between LLMs but not from a
  foreign-modality encoder (\cref{sec:res-rotation}).
  \item Controls that bound what the socket conveys: nearest-neighbour retrieval, homology splits
  and refits, lexical overlap, and the same DNA given to the host as text (\cref{sec:res-mechanism}).
  \item Beyond the benchmark, on two larger hosts: the socket follows the DNA string, not
  the protein, and the human-fitted plug transfers to non-human genes, beating retrieval where no human
  relative exists (\cref{sec:res-beyond}).
\end{enumerate}

%% file: sections/02-related-work.tex
\section{Related Work}
\label{sec:related}

\paragraph{Connecting biological encoders to LLMs.}
Most language systems for DNA, proteins and molecules train a connector, and often the LLM.
ChatNT \citep{Richard2024-ChatNT} trains a Perceiver resampler into a frozen Vicuna.
BioReason \citep{Fallahpour2025-BioReason} trains a linear projector from \evo{} into Qwen3 and
fine-tunes the host with supervised and reinforcement learning. Protein systems use trained
projectors or Q-formers \citep{Xiao2024-ProteinGPT,Liu2024-ProtT3,Liu2024-EvoLlama,Fei2025-Prot2TextV2},
gated cross-attention between frozen layers \citep{Zhou2025-Evolla}, or joint pretraining
\citep{Zhuo2024-ProtLLM,Wang2024-InstructProtein,Pei2023-BioT5,Xia2025-NatureLM}. Molecule models
follow BLIP-2 \citep{Li2023-BLIP2,Liu2023-MolCA,Li2024-3DMoLM}. Two works anchor the connector to the
host's text. GTA \citep{Honig2024-GTA} attends from DNA features to prototypes built from the token
embeddings of class names. BioVERSE \citep{Tsou2025-BioVERSE} aligns protein and cell embeddings (in one variant
contrastively) to pooled LLM text embeddings and then writes them into a placeholder slot. Both train
the connector and write at the input. The socket instead fits a ridge map, writes after an early
layer, and needs no gradient through the host.

\paragraph{Linear and closed-form maps.}
One linear map from a frozen encoder into a frozen LM's input space is enough for captioning
\citep{Merullo2022-LiMBeR}. Small projectors let LLMs learn in context from molecule, graph and brain
embeddings \citep{Zhuang2024-VectorICL}. Text-anchored connectors place visual tokens in the span of
the host's word embeddings \citep{Yin2024-SEA,Masry2025-AlignVLM,Liao2025-LangBridge}. These maps are
all trained through the host. Closed-form maps between models of one modality are well known.
Procrustes and least-squares maps between LLMs transfer residual states and steering vectors across
families
\citep{Zhou2026-CrossLingualAlignmentWithoutJointTraining,Huang2025-CrossModelPlatonicConcepts,Lee2025-SharedGeometryEmb2Emb}.
Ridge maps make one LLM's final states decodable by another \citep{Gorbett2026-Ventriloquist}, and a
learned projection of another LLM's states has no causal effect when injected into \mistral{}
\citep{CardenasPiepereit2026-CausalTransferActivationStates}. An
orthogonal map turns a sender's states into a receiver's input prefix \citep{Peng2026-StateBridge}.
The map in activation communication can be fitted by least squares
\citep{Ramesh2025-CommunicatingActivations}. Trained stitching maps carry features between models
\citep{Chen2025-StitchingLinearFeatures,Oozeer2025-TransferActivationInterventions}, and anchor-based
or closed-form translations relate latent spaces across models and modalities
\citep{Moschella2023-RelativeRepresentations,Maiorca2023-LatentSpaceTranslation,Jha2025-Vec2Vec}.
CAST \citep{Heyden2026-CAST} maps text descriptions in closed form into a classifier's weights, to
extend it to unseen classes without target-domain examples. Our cross-host plug transfer applies the
cross-LLM technique to modality plugs, and we claim it only as that. We have not found earlier work
with a closed-form map from a \emph{foreign-modality} encoder into a frozen LLM's \emph{own} text
representation at an intermediate layer. Late-layer ridge maps between LLMs fail below the last
layers \citep{Gorbett2026-Ventriloquist}; our maps target early-layer text states. Concurrently with and
independently of this work, \citet{Cheng2026-PlatosKey} transfer subject-token states between twelve
text-only LLMs with a closed-form rotation and one global scale, for factual recall. Their setting is the
cross-LLM case of our host-to-host map, which our first preprint (October 2026) already reported; we compare
their rotation with our ridge maps in both settings (\cref{sec:res-rotation}).

\paragraph{Writing hidden states.}
Replacing a token's hidden state at a chosen layer and continuing the forward pass is a standard
tool. Patchscopes and SelfIE use it to read representations
\citep{Ghandeharioun2024-Patchscopes,Chen2024-SelfIE}. Activation communication uses it between
agents \citep{Ramesh2025-CommunicatingActivations}. A compressed instruction can replace the state
after a placeholder at an early layer \citep{Ardoin2026-PromptCompressionActivationAggregation}.
Early-layer word states are read as words \citep{Kaplan2025-TokensToWords}, and entities are resolved
early and enriched before attributes are extracted
\citep{Geva2023-DissectingRecall,Biran2024-HoppingTooLate,Lad2024-StagesOfInference}. This fits our
finding that hosts stop reading slot writes after about a third of their blocks. The socket uses the
same operation. What is new is the source: a closed-form image of a foreign encoder in the host's
own text space.

\paragraph{Choosing the layer.}
Probe accuracy is a poor guide to where a steering vector acts \citep{Wang2026-ReadBestNotSteerBest}.
Grafting a vision encoder zero-shot depends on matching the host's early layers up to a logit-lens
transition point \citep{Yue2025-EncoderGrafting}. Our rule measures how well the host reads a written
slot, not where a property can be decoded. We tested it out of sample on a fourth host.

\paragraph{Shared representations and the global workspace.}
Middle layers share a text-dominated semantic space across languages and modalities
\citep{Wendler2024-LlamasWorkInEnglish,Wu2024-SemanticHubHypothesis}, and representations converge
across models \citep{Huh2024-PlatonicRepresentation}. In frozen LLMs, perceptual tokens stay in narrow
cones apart from text tokens \citep{Shukor2024-ImplicitMultimodalAlignment}; the socket instead writes
the host's own text representation.

\paragraph{Retrieval as an explanation.}
GenePT \citep{Chen2025-GenePT} represents a gene by a language-model embedding of its text
description and finds that this text alone carries much gene biology; the socket's target is the
host's own version of such a representation. For proteins, nearest-neighbour retrieval matches or
beats text-generating LLMs \citep{Wu2025-ProteinRetrievalOrLLM}. In LLM-based forecasting, removing the LLM often costs nothing
\citep{Tan2024-LMsUsefulForTimeSeries}. We therefore report a nearest-neighbour baseline and a
homology split as required controls. The new part is that a DNA plug refitted without any homologs
keeps most of its effect, and beats retrieval on non-human genes without human relatives.

%% file: sections/03-method.tex
\section{Method}
\label{sec:method}

\subsection{Hosts, encoders and task}
\label{sec:meth-setup}

\paragraph{Hosts.} We use four instruction-tuned LLMs from three developers, frozen throughout:
\qfour{} and \qtwentyseven{} \citep{Qwen2026-Qwen3.5}, which share a tokenizer and a hybrid layout
of Gated DeltaNet and attention blocks; \mistral{}-Instruct-v0.3 \citep{Jiang2023-Mistral7B}; and
OLMo-2-1124-7B-Instruct \citep{OLMo2024-OLMo2}, an ungated model from a third family. \olmo{} is the
out-of-sample host for the layer rule (\cref{tab:test}; details in \cref{tab:hosts}). Prompts use
each host's chat template with thinking turned off. Block $\ell$ is the output of transformer block
$\ell$ ($0$-based).

\paragraph{Encoders.} The frozen \evo{} 7B model \citep{Brixi2025-Evo2} encodes DNA. We take the MLP
output of block 20 (\texttt{blocks.20.mlp.l3}) over a gene's coding sequence, mean-pool it into one vector per gene, and z-score them
with training statistics. For the exploratory protein arm (\cref{sec:res-protein}), the frozen \esmc{}
600M model \citep{Candido2026-ESMC} encodes each gene's canonical protein (the translation of its
MANE Select coding sequence). We mean-pool its layer-30 activations in the same way.

\paragraph{\geneblind.} \geneblind{} is a benchmark of five-option multiple-choice questions about genes.
Its labels come from open resources (Reactome, Gene Ontology, UniProt and ClinVar
\citep{Landrum2018-ClinVar}). A question names the gene only through its DNA. The prompt has a DNA
preamble with $n$ placeholder positions (``slots''), then the question, five lettered options and an
answer instruction. There are four question families: \fpath{} (top-level Reactome pathway), \fdis{} (disease
group of pathogenic variants), \fblock{} (which process a compound must block to disable the product; two
hops) and \fshare{} (two sequences; which process both products share). The split assigns whole connected
components of a homology graph (edges: $\ge$50\% identity over $\ge$50\% of both proteins) to train,
development or test. Homology analyses use the components at 30\% identity (``homology groups''); an item's \emph{cluster} is the homology group of its first gene, and the bootstrap resamples clusters. Wrong options are chosen without reference to the answer's properties
(``answer-blind'' option sets). This removes most of the text-only prior of an earlier builder
(\cref{app:geneblind}). Of the 3,357 usable answer-blind development items, 2,972 are split by homology
into \devtune{} (1,508 items), used for curves and debugging, and \devconf{} (1,464 items, 377 clusters), used
only for confirmation under the frozen plan (385 items are dropped; \cref{app:geneblind}). Accuracy is
the share of items whose correct letter has the highest next-token logit among the five option
letters.

\subsection{The closed-form socket}
\label{sec:meth-socket}

\paragraph{Target.} For each training gene $g$ and host $M$, we run the host on three texts: the
official symbol, the protein name and the description (the first two sentences of the UniProt function text, at most
64 tokens; a gene without one uses its protein name, which then counts twice). We mean-pool the hidden
states after block $\ell$ over each text's tokens. The target is the mean of the three pooled
vectors, $t_g=\tfrac13\bigl(r^{\text{sym}}_g+r^{\text{name}}_g+r^{\text{desc}}_g\bigr)\in\mathbb{R}^{d_M}$.

\paragraph{Plug.} The plug is an affine map $P(x)=\mu_t+(x-\mu_x)W$ from the encoder feature $x_g$
to $t_g$. We fit it by ridge regression \citep{Hoerl1970-Ridge} on the 8,319 genes of usable
\geneblind{} training items that share no gene with the development set and have a protein name or description. The penalty is chosen from
nine values by 5-fold cross-validated $R^2$ on training genes only. The fit is closed form, in
float64, and deterministic. Ridge shrinks predictions toward the mean, so we rescale each output's
deviation from the target mean to the mean training deviation norm (the \emph{norm match}):
\begin{equation}
\hat h(x)=\mu_t+\bar\rho\,\frac{P(x)-\mu_t}{\lVert P(x)-\mu_t\rVert},\quad
\bar\rho=\frac1{|G|}\sum_{g\in G}\lVert t_g-\mu_t\rVert .
\end{equation}

\paragraph{Replace write.} The host runs normally up to block $\ell^*$. After block $\ell^*$, we
\emph{replace} the hidden state at every DNA slot by $\hat h(x)$, and the forward pass continues.
Nothing else changes, and no gradient passes through the host.

\paragraph{Modular map.} To feed host $Y$ with host $X$'s plug, we fit a second ridge map
$S_{X\to Y}$ from $X$'s target space at $\ell^*_X$ to $Y$'s at $\ell^*_Y$. It uses the same
training genes, but only their text representations. We then write $S_{X\to Y}(P_X(x))$ into
$Y$ with $Y$'s norm match, where $P_X$ is $X$'s plug before its norm match. Adding a host therefore needs only text, never DNA.

\subsection{Choosing the layer}
\label{sec:meth-layer}

The layer rule (the ``half-max rule'') asks whether the host can read a known text representation written at
the slots. For 500 training genes, it builds synthetic items in the \geneblind{} layout. Each item asks
for the gene's official symbol, with that symbol and four other genes' symbols as options. At layer
$\ell$, the rule replaces the slot states by the gene's pooled description representation or by zero
(sham). This gives $\text{gain}(\ell)=\text{acc}_\text{desc}(\ell)-\text{acc}_\text{sham}(\ell)$. Let
$S=\{\ell:\text{gain}(\ell)\ge\tfrac12\max_{\ell'}\text{gain}(\ell')\}$. The rule returns the probed
layer nearest to $(\min S+\max S)/2$. It uses no \geneblind{} item, label or DNA feature. We selected it
from 36 candidates with a criterion fixed in code before any probe ran (smallest mean relative regret
against the \devtune{} curves of 4B, 27B and \mistral{}). We then froze it and applied it to \olmo{}, and
committed the prediction before any plug scoring on that host (a sanity check that wrote the host's own
description at three layers ran first; \cref{app:deviations}). An earlier rule (the ``peak-margin rule'',
\cref{app:layer}) fixed the 4B and 27B layers before \devconf{}. A prediction succeeds if it is within one
block of the \devtune{} argmax or within 2\% of the maximal \devtune{} gap.

\subsection{Controls, statistics and the frozen plan}
\label{sec:meth-stats}

\paragraph{Controls.} We score every host on the same items with: (i) the plug applied to a
\emph{shuffled} sequence of the same gene (for DNA, the 5-let shuffle: the letters reordered so that every 5-letter word keeps its count, seeded per
item and slot); (ii) a zero write (\emph{sham}); (iii) five plugs refitted with the gene--target
pairing \emph{permuted}; (iv) a \emph{gene swap}, which gives each item the plug output of a gene from
another cluster; (v) a Gaussian \emph{random map}; (vi) a \emph{1-NN lookup}, which writes the target
of the nearest training gene in encoder space; (vii) a \emph{homology} split, and a plug refitted
after removing every training gene that shares a homology group with any development gene
(8,319$\to$6,291 genes); and (viii) a \emph{description-reading ceiling}, which writes the gene's own
text representation. The plan expects (i) and (iii)--(v) within $\pm2$\% of sham (\cref{app:controls}).

\paragraph{Endpoint and inference.} The gap is accuracy with real input minus accuracy with shuffled
input, pooled over all four families, in percent of the questions, at each host's frozen layer. Each gap has a 95\% percentile
interval from a cluster bootstrap over homology groups (2,000 resamples). The test is one-sided, with
Holm correction \citep{Holm1979-Holm} over hosts at $\alpha=0.025$ (it keeps the chance of any false
positive across the hosts at 2.5\%). The smallest effect size of
interest (SESOI) is 4\%.

\paragraph{The frozen plan.} We committed the analysis plan before any scoring on the answer-blind
development sets. We amended it twice, before any test item was built. The \mistral{}-layer amendment replaced the peak-margin rule
by the half-max rule for \mistral{} after the peak-margin rule failed there. The \olmo{} amendment added \olmo{} with its frozen half-max-rule
prediction. We picked \olmo{} by a fixed rule (no gated weights, a new model family, a named
fallback) and froze its layer before any plug scoring, but the \olmo{} amendment was approved only after \olmo{}'s
development results were known (\cref{app:deviations}). The test run is one pass over the four hosts with the development plugs (file hashes
checked, never refitted). Nothing may be tuned, added or dropped after test items are seen. \Devconf{} confirms the
item build and the layers on 4B and 27B. It does not confirm the plug design, which was chosen on
development genes that \devconf{} shares (\cref{sec:discussion}).

\subsection{Trained baseline}
\label{sec:meth-trained}

We re-score existing trained checkpoints of \arm{mlp}, a BioReason-style MLP adaptor from \evo{} into the
input embeddings at the slots, with the host and encoder frozen (\cref{app:trained}). Its gap is defined
as the socket's.

%% file: sections/04-results.tex
\section{Experiments and Results}
\label{sec:results}

All gaps are differences in accuracy, real minus shuffled, in percent of the questions (pooled over
all four families), with 95\%
cluster-bootstrap CIs. The primary result is the single run on the held-out test set
(\cref{sec:res-test}, \cref{tab:test}). ``\Devconf{}'' marks confirmatory development numbers under the
frozen plan (\cref{tab:main} in \cref{app:devb}). The layer curves, DNA as text, the rotation, the
protein plug and \cref{sec:res-beyond} are exploratory: selected or first seen on development data, or run
once outside the frozen plan.

\input{sections/05-test}

\subsection{A trained connector is unstable across seeds}
\label{sec:res-trained}

We re-scored existing trained checkpoints (\cref{app:trained}). On the test set, the
27B MLP adaptor's five seeds give $+24.3$, $+18.6$, $+9.5$, $+0.8$ and $+16.2$; on \devconf{} the same
seeds gave $+23.1$ to $+0.8$, while the socket is comparable to the MLP (\cref{sec:res-test}) with no
seed variance and no training.

\subsection{Where to plug}
\label{sec:res-layer}

\begin{figure*}[t]
\centering
\includegraphics[width=0.86\textwidth]{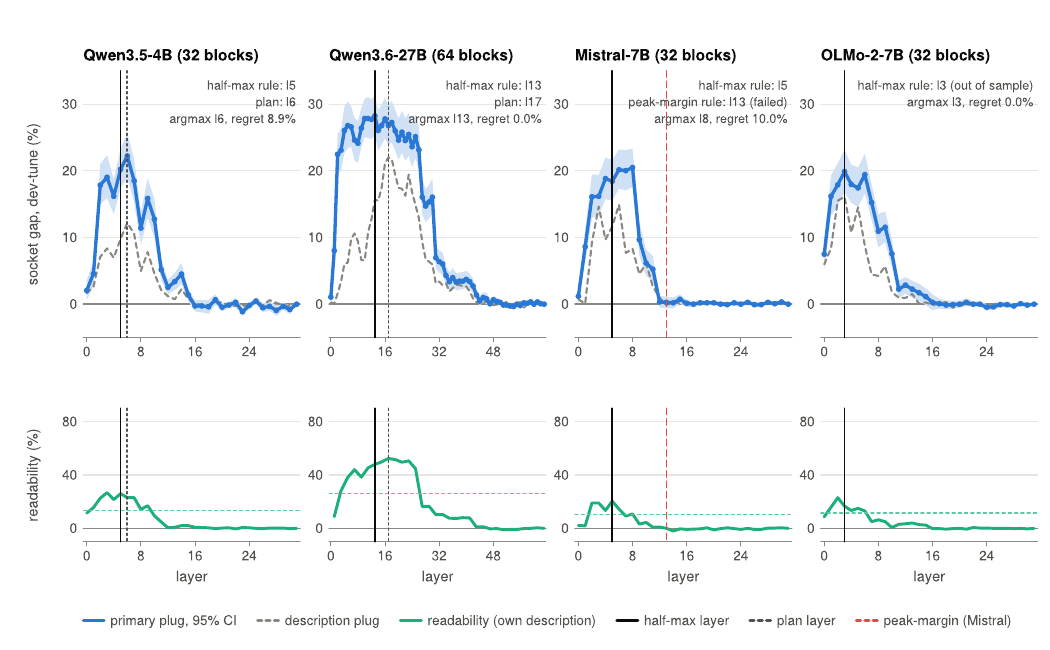}
\caption{\textbf{Socket gap versus layer} (\devtune{}). Top: primary plug with 95\% CI
(solid) and description plug (dashed). Bottom: the half-max rule's readability curve, with its half-max
line. The curve is the accuracy gain, on synthetic symbol questions, from writing the host's own
description at the slots instead of a zero write. Marks: half-max-rule prediction (out of sample for
\olmo{}), frozen test layer, and the peak-margin rule's failed \mistral{} prediction ($\ell=13$). Every block of
every host is scored; values and the layers added after the test run are in \cref{app:layer}.}
\label{fig:layers}
\end{figure*}

The socket works only in an early band of layers, and the band differs between hosts
(\cref{fig:layers}; \devtune{}). On \qfour{}, the gap peaks at $\ell=6$ ($+22.2$) and falls below $+6$
from $\ell=11$. \qtwentyseven{} has a broad plateau from $\ell=4$ to $\ell=26$ ($+23.1$ to $+28.2$)
falls to $+6.0$ by $\ell=33$ and stays below $+1$ from $\ell=44$ on. \mistral{} peaks at $\ell=6$--8 and gives nothing from $\ell=12$
on. \olmo{} has a plateau over $\ell=1$--6, with its argmax at $\ell=3$ ($+19.9$), and gives nothing
from about $\ell=11$. On these two hosts, the host also ignores its own description written at the
slots from the same depth, while the plug's cross-validated $R^2$ stays similar, so the limit is what the
host reads. The cross-form ``hub'' layers (4B $\ell=19$, 27B $\ell=35$) do not predict the working layer: the plug
gives $+0.7$ and $+3.4$ there.

The peak-margin rule patches description representations into a symbol question. It found the layer on both
Qwen hosts but failed on \mistral{}: its frozen prediction $\ell=13$ lies past the drop, 20.2\%
below the maximum. The half-max rule picks 4B $\ell=5$ and 27B $\ell=13$ (the argmax), which meet the
criterion in sample, and \mistral{} $\ell=5$, which misses it by 0.05\% (3 blocks from the
argmax). Out of sample, on \olmo{}, it predicted $\ell=3$ before any plug scoring, and the \devtune{}
argmax is $\ell=3$ (\cref{app:deviations}). The \olmo{} curve is a plateau, so this shows that the rule
lands in the usable band, not that it picks out one layer.

\subsection{Modularity: one plug, many hosts}
\label{sec:res-modularity}

A plug fitted for one host can feed another host through the text-to-text map $S_{X\to Y}$
(\cref{app:modularity}). On the test set we ran all twelve ordered pairs of the four hosts. Each
foreign plug gives within 0.9\% of the host's own plug, and every CI overlaps that of the own plug
(\cref{app:modularity}). So the DNA side needs one fit per encoder, and each new host needs only a
linear map learned from gene names and descriptions; we do not claim a universal socket.

\paragraph{Rotation or ridge} (\devconf{}, \cref{app:rotation}).
\label{sec:res-rotation}
A rotation with one global scale, fitted in closed form by orthogonal Procrustes
\citep{Schonemann1966-Procrustes} as in \citet{Cheng2026-PlatosKey}, can replace either of our maps. Between
two hosts' text spaces it matches ridge once each written vector gets the plug's norm match: rotation minus
ridge is $-0.3$ to $+0.5$\% on five hosts, and every CI includes zero. From \evo{} into a host it keeps
only 39--65\% of the ridge plug's gap (paired differences $-6.8$ to $-14.3$, every CI below zero). The text
spaces of two LLMs are close to a rotation of each other; a DNA encoder's space is not, which is why the
plug is a ridge regression.

\subsection{What the socket carries}
\label{sec:res-mechanism}

\paragraph{Identity.} A plug whose target is only the gene symbol gives $+9.8$ to $+19.4$ on test,
within 2\% of the description plug on three hosts (\cref{tab:test}). The gene swap removes the
effect. The socket tells the host which gene it is looking at.

\paragraph{Similar genes.} Writing the target of the nearest training gene (1-NN) gives $+18.4$,
$+22.0$, $+13.7$ and $+13.3$ on the test set, 72--84\% of the plug's gap. The plug is ahead of 1-NN
on all four hosts by $+3.0$ to $+5.2$\%, and every paired CI excludes zero (not corrected for the
four hosts). 1-NN depends on training homologs (items with one gain 7.1--11.4\% more, CIs exclude
zero); the plug much less ($-0.1$ to $+3.8$, CIs include zero), and refitted without any homologs it keeps
$+14.5$ to $+23.2$ (\cref{app:mechanism}). So the DNA plug does more than copy the nearest training gene,
though this is still a statement about gene identity.

\paragraph{Lexical overlap.} In 1,499 of 7,208 test items, a content word of the gene's description
appears in the correct option. These items gain more over sham (4B $+29.4$ vs $+19.8$; 27B $+31.8$ vs
$+25.0$). So part of the gain is lexical, but items without overlap still gain 16--25\%.

\paragraph{DNA as text} (\devconf{}). The hosts may have seen gene sequences in pretraining,
so we also pasted each gene's coding sequence into the prompt as letters, with no socket. Every host
stays at chance: real minus 5-let-shuffled letters is $+0.2$, $-0.3$, $+0.3$ and $+0.3$, and every CI
includes zero (\cref{app:dna-text}). With the gene symbol written instead, the same prompts give
40.9--58.7\%. The hosts cannot read a gene from its sequence; the socket's gain comes through the
encoder.

\subsection{Beyond the benchmark: mutant and non-human genes}
\label{sec:res-beyond}

\begin{figure*}[t]
\centering
\includegraphics[width=0.86\textwidth]{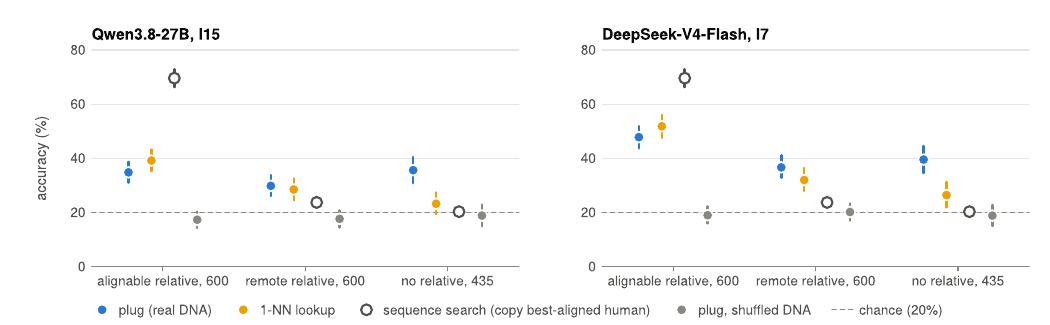}
\caption{\textbf{Non-human genes} by their relation to the human genes the plug was fitted on
(two larger hosts, one run each). Accuracy with 95\% CIs (bootstrap over the genes' homology
clusters); the number of questions follows each group. Sequence search copies the annotation of the
best-aligned human protein. Where a human relative aligns, search and 1-NN beat the plug; where none
exists, both fall to chance and the plug does not (\cref{app:larger}).}
\label{fig:nonhuman}
\end{figure*}

After the test run we applied the socket, unchanged, to two larger hosts outside the frozen plan:
\qbig{} and \dsflash{} (284B parameters; its four residual streams are written jointly, \cref{app:larger}).
The half-max rule chose blocks 15 and 7, and the plugs were fitted as for the four hosts. Each check below
is a single run.

\paragraph{The socket follows the DNA string.} We rewrote every \devconf{} gene's coding sequence and
wrote the rewrite through the same plug. A frameshift (one letter deleted, protein nonsense after residue
10) and an early stop codon leave accuracy unchanged (within 0.3\% of real DNA on both hosts).
Synonymous recoding, which keeps the protein but changes 42\% of the letters, costs 4.6\% \ci{2.8}{6.5}
and 2.8\% \ci{0.8}{4.7}. Reversed DNA is at the shuffled level. So the plug reads what \evo{} makes
of the sequence, not the encoded protein.

\paragraph{The plug transfers to non-human genes.} We built 1,635 five-option questions in
\geneblind{}'s answer-blind format about 1,557 genes of yeast, \textit{Arabidopsis}, \textit{E.~coli} and
an archaeon (GO-slim process, or the reaction subclass of enzymes) and sorted the genes by their best
alignment to the 18,758 human proteins of the study: an alignable relative ($E\le10^{-3}$), a remote one
(no alignment, but a shared InterPro superfamily, domain or family), or none. With the human-fitted plug,
real DNA beats shuffled DNA by $+15.3$ \ci{12.9}{17.7} (\qbig{}) and $+22.1$ \ci{19.1}{25.1}
(\dsflash{}), while the DNA of another gene of the same organism stays near the shuffled level.
\Cref{fig:nonhuman} shows a crossover. Where a human relative aligns, copying its annotation gives
69.6\% and 1-NN beats the plug by about 4\%. Where no relative exists, both fall to chance, while the
plug keeps 36--40\% and beats 1-NN by $+12.4$ \ci{7.3}{17.7} and $+13.1$ \ci{7.7}{18.9}. In the plug's
input space these genes lie far from all human genes (median cosine to the nearest human gene 0.38,
against 0.65 between unrelated human genes). The plug thus generalises beyond retrieval of a similar
human gene; it still conveys coarse function the host can name, not knowledge the host lacks.

\subsection{Second modality: protein}
\label{sec:res-protein}

A ridge plug from \esmc{} protein features, fitted exactly like the DNA plug and scored against a residue
shuffle, gives $+19.0$ to $+25.8$\% on \devconf{} in the same frozen sockets (developed on
\devtune{}, never run on test), with null controls and cross-host transfer as for DNA. But it is about as
good as 1-NN retrieval in \esmc{} space (paired difference $+0.2$ to $+2.9$, every CI includes zero), so we
read it as accepting a second modality at the level of nearest-homolog retrieval
(\cref{tab:protein}, \cref{app:protein}).

%% file: sections/05-test.tex
\begin{table*}[t]
\centering
\small
\setlength{\tabcolsep}{5pt}
\caption{\textbf{Main result: the single run on the held-out test set} (7,208 questions about genes never used for
fitting). Accuracy: share of questions answered correctly (chance 20\%). \emph{Gap}: accuracy with the gene's
real DNA minus accuracy with a shuffled copy of it, in \% of the questions, with 95\% confidence interval.
Empty: nothing written (zeros). Significant; above 4\%: passes the test corrected for four models, and the whole
interval lies above 4\%. Nearest training gene: the most similar training gene's target written instead of the
plug's output. Refitted without relatives: the map fitted again without any training gene related to a test
gene. All rows: \cref{tab:test-full}.}
\label{tab:test}
\begin{tabular}{@{}lcccc@{}}
\toprule
 & \qfour{} & \qtwentyseven{} & \mistral{} & \olmo{} \\
\midrule
Block written after / blocks          & 6 / 32 & 17 / 64 & 5 / 32 & 3 / 32 \\
How the block was chosen                    & peak-margin & peak-margin & half-max (in sample) & half-max (held out) \\
Accuracy: empty / real / shuffled   & 20.0 / 41.8 / 19.6 & 19.6 / 46.0 / 19.9 & 19.8 / 36.8 / 20.1 & 19.6 / 38.7 / 20.2 \\
\textbf{Gap (real $-$ shuffled)}                  & \textbf{+22.2} \ci{19.8}{25.0} & \textbf{+26.1} \ci{24.3}{27.9} & \textbf{+16.7} \ci{14.2}{19.7} & \textbf{+18.4} \ci{16.2}{21.0} \\
Significant; above 4\%             & reject, above & reject, above & reject, above & reject, above \\
Gap without \fdis{} questions                    & +23.5 \ci{20.8}{26.5} & +27.5 \ci{25.7}{29.3} & +17.7 \ci{15.2}{20.9} & +19.6 \ci{17.2}{22.3} \\
Real $-$ another gene's DNA                 & +21.9 \ci{19.4}{24.5} & +26.1 \ci{23.7}{28.8} & +16.9 \ci{14.4}{20.2} & +19.6 \ci{17.4}{22.2} \\
Gap, nearest training gene                      & +18.4 \ci{15.8}{21.6} & +22.0 \ci{19.1}{25.4} & +13.7 \ci{11.0}{17.2} & +13.3 \ci{10.8}{16.3} \\
Plug $-$ nearest training gene        & +3.8 \ci{2.3}{5.2}    & +4.1 \ci{1.6}{6.3}    & +3.0 \ci{1.4}{4.5}    & +5.2 \ci{3.8}{6.5} \\
Gap, refitted without relatives            & +19.3 \ci{17.8}{20.9} & +23.2 \ci{21.8}{24.6} & +14.5 \ci{12.6}{16.6} & +16.1 \ci{14.5}{17.8} \\
Gap, aimed at symbol / description    & +12.8 / +11.8 & +19.4 / +21.0 & +9.9 / +10.9 & +9.8 / +14.2 \\
Trained MLP adaptor (mean)        & +17.3 \ci{4.1}{30.5} & +13.9 \ci{2.7}{25.1} & --- & --- \\
Gap on development data (\cref{tab:main})   & +23.8 & +26.0 & +17.2 & +20.7 \\
\bottomrule
\end{tabular}
\end{table*}

\subsection{The socket on the held-out test set}
\label{sec:res-test}

\paragraph{Primary result.} The single test run follows the frozen plan with its two amendments (\mistral{} layer, \olmo{} added).
It scores 7,208 answer-blind \geneblind{} test items (2,471 genes, 1,535 clusters) once. No item was
dropped for a missing coding sequence or feature. Every test slot has its own 5-let shuffle, the
plugs are the development plugs (file hashes checked, never refitted), and each host is scored at its
frozen layer. On all four hosts, real DNA beats shuffled DNA (\cref{tab:test}): $+22.2$
\ci{+19.8}{+25.0} on \qfour{}, $+26.1$ \ci{+24.3}{+27.9} on \qtwentyseven{}, $+16.7$
\ci{+14.2}{+19.7} on \mistral{} and $+18.4$ \ci{+16.2}{+21.0} on \olmo{}. All four reject $H_0$ after
Holm over four hosts ($p<0.0005$), and all four lower bounds exceed the SESOI. Without the weak \fdis{}
family the gaps are 1.0--1.4\% larger. The test gaps are within 2.3\% of \devconf{}.

\paragraph{Controls.} Every null control is within the plan's $\pm2$\% of sham on every host.
Shuffled DNA minus sham is $-0.4$ to $+0.6$; the five permuted-gene plugs give $-1.1$ to $+0.9$
against sham, the random map $-1.1$ to $+0.1$, and the gene swap $-0.6$ to $+0.2$. Real$-$swap is
$+16.9$ to $+26.1$. We make no claim that replacing beats adding
(\cref{app:controls}).

\paragraph{Trained connector.} The socket's 4B gap lies inside the 4B MLP adaptor's seed interval
(three seeds, 26\% wide), so by the plan's rule it is \emph{comparable}. On 27B it lies 1.0\% above the MLP's upper bound, which the plan's rule labels ``$\geq$ MLP''. This margin is
fragile: leaving out any one of four of the five seeds puts the socket inside the interval, and on
the earlier development pack the margin did not survive the sensitivity analysis without \fdis{}. We therefore read the socket as comparable
to the MLP on both hosts. One 27B MLP seed barely uses the DNA ($+0.8$). The run took about an hour on four GPUs per host group (\cref{app:repro}).

%% file: sections/06-discussion.tex
\section{Discussion and Limitations}
\label{sec:discussion}

\paragraph{A baseline that trained connectors should beat.} One ridge fit, written once into an early
layer (its cost includes one pass of the host over the gene texts to make the targets, 1--2.5\,min per
host; \cref{app:repro}), gives a frozen LLM about as much usable gene information on the \geneblind{} test set as a
trained MLP adaptor does on average, with no failed seeds and no gradient through the host. Any
DNA--LLM connector that claims to deliver sequence information should report its gain over this socket,
on items held out by homology, with gene-swap and nearest-neighbour controls.

\paragraph{What reaches the host is identity.} The controls, the symbol plug, the 1-NN lookup and the
homology analyses agree. The socket tells the host which gene, or which kind of gene, it is looking
at, and the host answers from what it already knows. On the test set the DNA plug beats nearest-neighbour
retrieval on all four hosts, by 3.0--5.2\%, and a plug refitted without any homologs keeps most
of its effect, so it does more than copy a neighbour. On non-human genes without a detectable human
relative this lead grows to 12--13\%, so the plug interpolates in the encoder's space rather than
retrieving a human gene. The identity comes through the encoder: the hosts cannot read the sequence as
text, and the socket tracks the DNA string, not the protein. But nothing here shows the host using
sequence-level information its text knowledge lacks, and a large benchmark gain is compatible with a
learned lookup of gene identity, a warning for trained connectors too.

\paragraph{The hub is open, early, and only nearly Platonic.} Latent transfer works: the hub accepts
information from a model of another modality. But its entrance is early, at 9--27\% of depth, well before
the cross-form hub; two hosts stop reading a written slot state after about a third of their blocks, even their own
text representation. Slot readability passed its one out-of-sample test, and its layers also worked on the
two larger hosts. The rotation result qualifies the Platonic picture: the text spaces of two LLMs are a
rotation of each other up to length, as convergence predicts, but a DNA model's space reaches an LLM's
only through a general linear map.

\paragraph{Limitations.}
(i)~\emph{Plug design on shared genes:} the target, norm match and replace rule were chosen among about
25 variants on development genes that \devconf{} shares; only the test run, a single run on one
benchmark, is clean for them.
(ii)~\emph{Layer rule:} selected in sample on three hosts, one out-of-sample success on a plateau host;
the peak-margin rule's out-of-sample prediction failed.
(iii)~\emph{Scope:} one encoder layer, one benchmark family; sequence-level tasks (e.g.\ variant effects)
and retrieval beyond 1-NN are untested. The protein arm, DNA as text and the mutant and non-human checks
are exploratory; the last two ran once, outside the frozen plan, on two hosts (one converted from FP8/FP4
to bf16), with coarse non-human questions and no test of simple sequence properties as an explanation.
(iv)~\emph{Lexical share:} part of the gain comes from description words in the correct option.
(v)~The \fdis{} family is weak on every host ($+4.3$ to $+9.0$ on test). Its pre-planned retention rule, which
has not run, may drop it; we report the gap without \fdis{} as a sensitivity analysis.
(vi)~\emph{The trained MLP} checkpoints are existing ones from an earlier option builder, only re-scored and
descriptive; none exists for \mistral{} or \olmo{}.
(vii)~Deviations from the plan are listed in \cref{app:deviations}.

\paragraph{Conclusion.} We map a frozen biological encoder in closed form into a frozen LLM's own
text representation, and write the result at an early layer chosen without task items. In a single
run on a held-out test set, this lets four LLMs from three developers answer gene-level questions
about sequences whose names they are not told, at no training cost: $+16.7$ to $+26.1$\% over
shuffled DNA, with every host surviving Holm correction and every null control at the sham level.
The plug moves between hosts through a text-only linear map, within 1\% of the host's own plug.
It transfers gene identity, beyond 1-NN lookup and, in exploratory checks, beyond the species it was fitted on, but not as sequence-level knowledge the host lacks. We offer it as the baseline against which trained DNA--LLM connectors should be measured, and as a
probe of the host's hidden space; the next test is an encoder whose input the host has never seen
described as text.

%% file: sections/07-impact.tex
This paper presents a way to connect frozen biological sequence encoders to frozen language models
without training. It also presents controls that show what such connections convey. We see two
consequences. First, the method lowers the compute needed to build DNA--LLM systems: one pass of
the host over the gene texts, one ridge fit per encoder and a text-only map per host, instead of training a connector through models with
billions of parameters. This makes such systems easier to build and to audit. Second, our results
show that the gain comes mainly from recognising which gene a sequence belongs to, not from new
sequence-level knowledge. Benchmark gains of this method and of trained systems should therefore not
be read as evidence that a model can interpret new or variant sequences. Nothing here supports
clinical use. The method adds no capability the host lacks: it lets a host retrieve what it already
knows about known human genes. All data are public (Reactome, Gene Ontology, UniProt, ClinVar and
reference human coding sequences); we used no individual-level human data.

%% file: sections/A-appendix.tex
\section{\geneblind{} details}
\label{app:geneblind}

\paragraph{Gene universe and items.} The gene universe is the set of human protein-coding genes
with a MANE Select transcript whose coding sequence is at most 8,192\,bp. The number of slots $n$ in
a prompt depends on the length of the coding sequence. \fpath{} asks in which top-level Reactome pathway
the product takes part. \fdis{} asks with which disease group the gene's pathogenic variants are most
often associated (ClinVar). A \fdis{} answer is any MONDO disease area with at least two submissions summed
over the gene's pathogenic or likely pathogenic ($\ge$2-star) variants, not only the most frequent
area. Each such (gene, area) pair can give an item, with at most two \fdis{} items per gene, and wrong
options exclude every area of the gene's pathogenic variants. \fblock{} asks which process a compound must block to disable the product (a
two-hop question with a text premise). \fshare{} gives two sequences and asks which process both products
share. The split unit is a connected component of the homology graph whose edges join two
proteins with $\ge$50\% identity over $\ge$50\% of both sequences ($E\le10^{-3}$, exact
Smith--Waterman). Whole components are assigned to train, development or test (70/10/20), stratified
by the component's majority top-level Reactome pathway (Reactome's ``Disease'' pathway excluded). So no test gene has a homolog in training
above this threshold. Homology analyses use the components of the same graph at 30\% identity
(``homology groups''), which are coarser.

\paragraph{Answer-blind options.} The first option builder left a text-only prior. A model that saw
only the options reached 28--31\% (\fdis{} about 45\%), against a chance level of 20\%. Balancing the
ranks of the answer's label frequency and co-annotation (``hub'') lowered this to about 22\% under the
pre-registered text-only model (the ``frequency-balanced'' sets; stronger pairwise models still reached about 30\% pooled). All results in this paper use the answer-blind option
sets (one rule for \fpath{}--\fblock{} and a two-gene variant for \fshare{}; option seed 0). In these sets, wrong options are chosen
without reference to the answer's properties, and the text-only prior is at chance. We fitted ten
option-only predictors on training items: answer-label frequency and rate, co-annotation PMI, and
conditional, directional and pairwise logit models over the option labels. \fpath{}--\fblock{} have an eleventh,
a consistency heuristic. On the option-seed-0 development items, these predictors reach
18.5--20.8\% on \fpath{} (1,319 items), 16.0--21.4\% on \fdis{} (262), 19.0--20.9\% on \fblock{} (975),
16.9--22.0\% on \fshare{} (801) and 18.4--20.7\% pooled (3,357), against 20\% chance. Option seeds 1
and 2 stay within 15--25\% on every family.

\paragraph{Development halves and scoring.} Of the 3,357 usable answer-blind development items, we
dropped 4 because a DNA slot had no 5-let shuffle. We split the remaining 3,353 into two halves by
homology group and dropped the 381 items whose genes fall in both halves
(\cref{app:deviations}). Of the remaining 2,972 items, \devtune{} (1,508) was available for curves and
debugging. \Devconf{} (1,464 items, 377 clusters) was used only for confirmation under the frozen plan. We
committed the split manifest before any scoring (\cref{app:deviations}). Accuracy is read from the
next-token logits at the answer position, with the prompt given up to that position. The probability mass on the five letters is at least 0.70 on every host with
a zero write, and above 0.99 on most hosts and arms.

\section{Hosts, layers and the layer rules}
\label{app:layer}

\begin{table}[h]
\centering
\small
\caption{\textbf{The four language models and where the DNA is written.} $L$: number of blocks; $\ell^*$: the block
the DNA is written after on the test set, fixed in the analysis plan before the test, and its relative depth
$\ell^*/L$. The descriptive layer is a neighbouring block, also scored on the test set but without a test.}
\label{tab:hosts}
\begin{tabular}{@{}lccccl@{}}
\toprule
Host & Blocks $L$ & Test layer $\ell^*$ & Rel.\ depth & Descriptive layer & Source of $\ell^*$ \\
\midrule
\qfour{}        & 32 & 6  & 0.19 & 5 (half-max rule) & peak-margin rule, frozen before \devconf{} \\
\qtwentyseven{} & 64 & 17 & 0.27 & 13 (half-max rule) & peak-margin rule, frozen before \devconf{} \\
\mistral{}      & 32 & 5  & 0.16 & 8 (\devtune{} argmax) & half-max rule (in sample) \\
\olmo{}         & 32 & 3  & 0.09 & --- & half-max rule, frozen out of sample \\
\bottomrule
\end{tabular}
\end{table}

\paragraph{Revisions.} \evo{} 7B revision \texttt{074097e9} (MLP output of block 20, sub-module \texttt{blocks.20.mlp.l3},
before the residual sum); OLMo-2-1124-7B-Instruct weights revision \texttt{470b1fba}.

\paragraph{Peak-margin rule ($m_\text{sym}$).} At layer $\ell$, patch the host's pooled description
representation of a gene into a prompt that asks for the gene's official symbol. Measure the
double-centred margin of the correct symbol, and take $\ell^*=\arg\max_\ell m_\text{sym}(\ell)$. We
picked this rule from about 26 candidate statistics by its agreement with the 4B and 27B development
curves. It fixed $\ell^*=6$ (4B) and $\ell^*=17$ (27B) before \devconf{}. Its first out-of-sample
prediction, \mistral{} $\ell^*=13$, was committed before any \mistral{} scoring, and it failed.

\paragraph{Half-max rule details.} The 500 training genes are drawn with seed 20261011; ties go to the
lower layer. The in-sample picks are 4B $\ell=5$ ($+20.2$ vs the maximum $+22.2$), 27B $\ell=13$ (the
argmax) and \mistral{} $\ell=5$ ($+18.4$ vs $+20.5$; 3 blocks from the argmax, and 0.05\% short
of the 2\% criterion). For \olmo{}, it predicted $\ell=3$ (half-max band $\ell=1$--6). On the
older development set, moving the description plug one or two blocks away from the chosen layer cost
2.7--5.3\% (4B $\ell=7$, 27B $\ell=15$). So on hosts with a narrow band, the choice matters. On
\mistral{} the plug's cross-validated $R^2$ is 0.29 at $\ell=13$ against 0.33 at $\ell=7$, although the gap
falls to zero, so the limit is what the host reads, not what the plug predicts. On \olmo{} the gap at
$\ell=6$ is within 0.5\% of the maximum. \olmo{}'s prediction was committed after a description
sanity check at $\ell=4$, 12 and 20, and \olmo{} was kept in the test plan after its development results
were known (\cref{app:deviations}).

\paragraph{Full layer curves (\devtune{}, primary plug).} Every block of every host. The layers marked
``added'' were scored after the test run with the same command and items, to complete the curves (4B
$\ell0$ and $\ell17$--31, 27B every layer but the odd $\ell5$--33, \mistral{} and \olmo{} $\ell31$); the
added 27B layers ran on an A100, the others on the GPU type of the recorded runs. Layers scored both times
agree exactly on the same GPU type (4B $\ell16$, \mistral{} and \olmo{} $\ell30$) and within 0.6\%
across GPU types (27B $\ell5$ and $\ell33$); the figure and the list keep the recorded value. At the last
block the gap is 0 by construction: a state written after the last block reaches no later position.
\qfour{}: $\ell0$ $+2.1$ (added), $\ell1$ $+4.6$, $\ell2$ $+17.8$, $\ell3$ $+19.0$, $\ell4$ $+16.2$, $\ell5$
$+20.2$, $\ell6$ $+22.2$, $\ell7$ $+18.5$, $\ell8$ $+11.4$, $\ell9$ $+15.8$, $\ell10$ $+12.7$, $\ell11$ $+5.2$,
$\ell12$ $+2.5$, $\ell13$--16 $\le+4.5$, $\ell17$--31 (added) within $[-1.1,+0.7]$.
\qtwentyseven{}: $\ell0$ $+1.1$, $\ell1$ $+8.0$, $\ell2$ $+22.5$, $\ell3$ $+23.1$, $\ell4$ $+26.1$ (added);
$\ell5$ $+26.8$; $\ell6$ $+26.6$ (added); $\ell7$ $+24.6$; $\ell8$ $+24.1$ (added); $\ell9$ $+26.4$; $\ell10$
$+27.9$ (added); $\ell11$ $+27.9$; $\ell12$ $+27.7$ (added); $\ell13$ $+28.2$; $\ell14$ $+26.1$ (added); $\ell15$
$+26.8$; $\ell16$ $+27.8$ (added); $\ell17$ $+26.8$; $\ell18$ $+27.2$ (added); $\ell19$ $+26.0$; $\ell20$ $+24.6$
(added); $\ell21$ $+25.8$; $\ell22$ $+24.5$ (added); $\ell23$ $+25.5$; $\ell24$ $+23.6$ (added); $\ell25$ $+25.1$;
$\ell26$ $+23.1$ (added); $\ell27$ $+16.1$; $\ell28$ $+14.7$ (added); $\ell29$ $+15.3$; $\ell30$ $+16.1$ (added);
$\ell31$ $+7.0$; $\ell32$ $+6.4$ (added); $\ell33$ $+6.0$; $\ell34$--43 (added) $+1.4$ to $+4.3$; $\ell44$--63
(added) within $[-0.3,+1.0]$.
\mistral{}: $\ell0$ $+1.2$, $\ell1$ $+8.6$, $\ell2$ $+16.1$, $\ell3$ $+16.2$, $\ell4$ $+18.8$, $\ell5$
$+18.4$, $\ell6$ $+20.2$, $\ell7$ $+20.0$, $\ell8$ $+20.5$, $\ell9$ $+9.7$, $\ell10$ $+6.2$, $\ell11$
$+5.2$, $\ell12$--31 within $[-0.1,+0.7]$ ($\ell13$: $+0.3$ \ci{-0.6}{+1.1}; $\ell31$ added).
\olmo{}: $\ell0$ $+7.5$, $\ell1$ $+16.2$, $\ell2$ $+17.9$, $\ell3$ $+19.9$, $\ell4$ $+18.0$, $\ell5$
$+17.4$, $\ell6$ $+19.4$, $\ell7$ $+15.2$, $\ell8$ $+10.9$, $\ell9$ $+11.5$, $\ell10$ $+7.6$, $\ell11$
$+2.2$, $\ell12$ $+2.9$, $\ell13$--31 within $[-0.5,+2.2]$ ($\ell31$ added).

\paragraph{\mistral{} at the \devtune{} argmax.} At $\ell=8$, \devconf{} gives $+20.7$ \ci{+16.6}{+25.1}
(descriptive, because the layer was chosen on \devtune{}). This is 3.5\% above the half-max-rule layer.

\section{Controls in full}
\label{app:controls}

\begin{enumerate}
  \item \textbf{Shuffled DNA} (part of the endpoint): the plug applied to the \evo{} features of a
  5-let shuffle of the same gene's coding sequence. The shuffle is an Euler-path $k$-let shuffle with
  $k=5$, which keeps every 5-mer count. It is seeded per item and slot, and paired per item.
  \item \textbf{Sham}: a zero vector written with the replace rule.
  \item \textbf{Gene-permutation null $\times5$}: the plug refitted with the training gene--target
  pairing permuted (seeds 1--5), with the same norm match.
  \item \textbf{Gene swap}: each item receives the plug output of another dev gene from a different
  cluster (a seeded derangement); we report real$-$swap.
  \item \textbf{Random map}: a Gaussian matrix in place of the ridge map, with the norm match.
  \item \textbf{1-NN lookup}: the target of the nearest training gene by cosine on the z-scored
  encoder features, with the same norm match; we report the paired plug$-$1-NN difference.
  \item \textbf{Homology}: the gap on items with and without a training gene in the same homology
  group, and a refit of the plug (and 1-NN) after removing every training gene that shares a homology
  group with any development gene.
  \item \textbf{Description-reading ceiling}: the item gene's own description (or mean target)
  representation, written in place of the plug. This measures how well the host reads its own text
  at the slots. It is not an upper bound on the information in the DNA. We split it by whether a
  content word of the gene's description appears in the correct option.
\end{enumerate}

\paragraph{Test results.} On the test set every control is within $\pm2$\% of sham on every host
(\cref{tab:test-full}). As gaps against shuffled DNA, the permuted plugs give $-1.3$ to $+1.6$ and the
random map $-0.5$ to $+0.3$. Four control gap CIs exclude zero, all small: 4B perm.\ 3 $+1.3$
\ci{+0.1}{+2.4}, \olmo{} perm.\ 3 $+1.6$ \ci{+0.4}{+2.8}, \mistral{} perm.\ 2 $-1.3$ \ci{-2.3}{-0.3}
and the \olmo{} swap gap $-1.2$ \ci{-2.4}{-0.0}. Against sham, two more exclude zero, both below sham:
\mistral{} perm.\ 2 $-1.1$ \ci{-2.1}{-0.0} and the \mistral{} random map $-1.1$ \ci{-2.1}{-0.1}.

\paragraph{\Devconf{} results.} On 4B and 27B, shuffled DNA minus sham is $+0.4$ \ci{-2.1}{+2.7} and $-0.8$
\ci{-3.1}{+1.5}. The permuted plugs give gaps of $-1.8$ to $+1.1$ and $-1.9$ to $+0.9$, and the random
map gives $-0.3$ and $+1.6$. On 27B, three of the five permutation arms lie 2.1--3.0\%
\emph{below} sham and a fourth lies 2.0 below, all with CIs that include zero. So the plan's $\pm2$ expectation misses, but on the
conservative side. On \mistral{}: shuffled $-$ sham $+1.4$, permutations $-1.4$ to $+1.3$, random map
$-1.2$, real$-$swap $+17.6$. On \olmo{}: shuffled $-$ sham $-2.0$ \ci{-5.1}{+0.9} (30 items, 2.05\%, just outside the band), permutations $-0.4$ to $+1.7$, random map $-0.2$,
real$-$swap $+21.2$. As on 27B, one permutation and the swap arm lie 2.5--2.9\% below sham.

\paragraph{Replace versus add.} On 27B $\ell=17$, we add the plug to the slot state
($h\leftarrow h+a\,\hat h$) instead of replacing it. This gives $+0.5$, $+1.2$, $+3.6$, $+7.5$ and
$+8.9$ at $\times1$, $\times2$, $\times4$, $\times8$ and $\times16$ the base amplitude (\devconf{}), and
the gap is still rising. On the test set the same sweep gives $+0.3$, $+0.9$, $+2.5$, $+6.0$
\ci{+4.8}{+7.3} and $+7.5$ \ci{+6.1}{+8.9}, against $+26.1$ for replacing. The gain per doubling
shrinks, but the $\times8$ and $\times16$ CIs overlap, so the plan's condition for a ``replace beats
add'' claim (a plateau below replace by $\times16$) is not met.

\begin{figure}[h]
\centering
\includegraphics[width=0.85\linewidth]{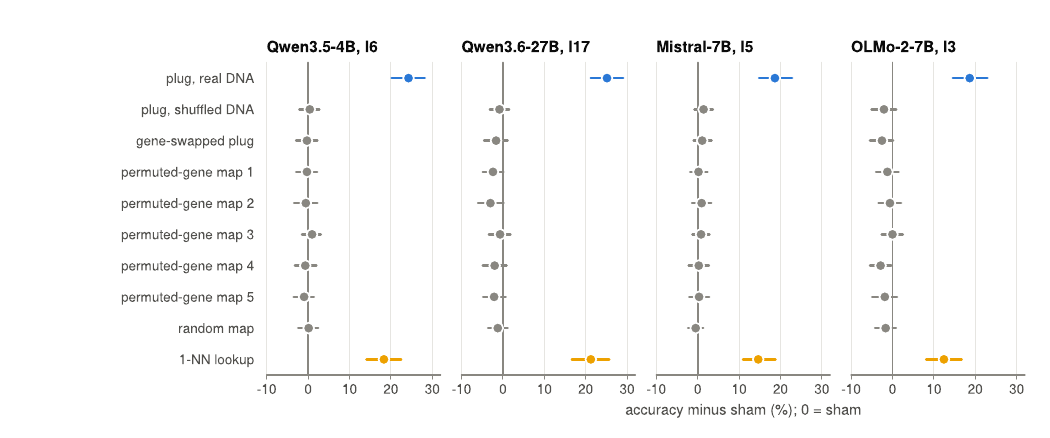}
\caption{\textbf{Controls on \devconf{}}, as accuracy minus the sham (zero-write) accuracy, for the four
hosts at their frozen layers. The plug on real DNA is far above sham. The plug on shuffled DNA, the
gene-swapped plug, five permuted-gene maps and a random map are at or below sham. The 1-NN lookup
recovers most, but not all, of the plug's effect.}
\label{fig:controls}
\end{figure}

\section{Trained connector}
\label{app:trained}

\paragraph{Arm.} \arm{mlp} keeps the host and the encoder frozen: a trained MLP adaptor from the \evo{}
features into the host's input embeddings at the slots, as in BioReason \citep{Fallahpour2025-BioReason}.
It was trained on two sets: the description stage on gene descriptions (\textsc{GeneDesc}, 45{,}412 examples)
and the question stage on \geneblind{} training items (the order in which they ran is given below). During
training, a \arm{null-dna} arm (the same pipeline without the DNA input) served as a reference. It is not
used here: as for the socket, the MLP's gap is its accuracy with real minus 5-let-shuffled DNA, both passed
through the same trained connector.

\paragraph{Training data.} The checkpoints predate the answer-blind option sets. The question stage used
\geneblind{} training items built with an earlier option builder, whose options still carried a
text-only prior (\cref{app:geneblind}); we only re-score the checkpoints on answer-blind items. The
socket's plug saw no items at all. This shift can lower the MLP's absolute accuracy, but its gap remains a
valid measure of DNA use.

\paragraph{Configuration} (the runs re-scored here). A trunk on the \evo{} features,
$\text{Linear}(8192,2048)\to\text{GELU}\to\text{Linear}(2048,2048)\to\text{GELU}$, and a
$\text{Linear}(2048,d)$ head whose output is a soft token at the slots, where $d$ is the host's hidden
width. The input has $d_\text{enc}=8{,}192$: the block-20 MLP-output features of the forward and
reverse-complement strands, concatenated. Training minimises the cross-entropy of the answer token. It uses
AdamW (learning rate $10^{-3}$, weight decay 0.01), a linear warm-up over 5\% of the steps, a cosine decay
to zero, and a global batch of 16 sequences (micro-batch 4). The run has 7,395 steps over 45,408 of the
45,412 \textsc{GeneDesc} examples (the description stage, 2,838 steps per pass, 8.4M tokens) and 26,216
\geneblind{} training items (the question stage, 1,639 steps per pass). The sampler orders each epoch as the
description stage, then the question stage. One epoch is 4,477 steps, so the run was the description stage
(2,838 steps), the question stage (1,639), the description stage again (2,838) and 80 steps of the question
stage: 1,719 question-stage steps, about 1.05 passes over its items, and not the 4,557 question-stage steps
we intended (\cref{app:deviations}). Seeds: 1--5 for 27B, 1--3 for 4B. The 27B runs use FSDP2 over two
H100s with activation checkpointing; the 4B runs use one H100.

\paragraph{Test results.} On 27B, the MLP's five seeds give $+24.3$, $+18.6$, $+9.5$, $+0.8$ and
$+16.2$ (mean $+13.9$, $t$-CI \ci{+2.7}{+25.1}). On 4B, they give $+11.2$, $+20.1$ and $+20.6$ (mean
$+17.3$ \ci{+4.1}{+30.5}). The socket gap is inside the 4B MLP interval (``comparable'') and above the
upper bound of the 27B MLP interval, which is the plan's condition for a ``$\geq$ MLP'' statement.
Both readings are weak. The 4B interval comes from three seeds and spans 26\%. On 27B the margin
is 1.0\%: leaving out seed 2, 3, 4 or 5 gives upper bounds of $+28.6$, $+31.0$, $+27.0$ and
$+29.8$, all above the socket's $+26.1$, and only leaving out seed 1 keeps ``$\geq$''. On \devconf{} the
same statement held by 0.2\% (0.4 on the earlier development pack, where it failed the sensitivity analysis
without \fdis{}; \cref{app:deviations}); on test the MLP was not re-analysed without \fdis{}. We therefore
call the socket comparable to the MLP on both hosts. The socket minus the MLP seed mean is $+4.9$ (4B) and
$+12.2$ (27B).

\paragraph{\Devconf{} results.} On 27B, the MLP's five seeds give $+23.1$, $+21.9$, $+11.2$, $+0.8$ and $+15.6$
(mean $+14.5$, $t$-CI \ci{+3.3}{+25.8}). On the older development set, the same seeds gave $+25.9$ to
$+1.5$, and a sixth seed gave $-0.3$. On 4B, three MLP seeds give a mean of $+19.1$ \ci{+3.8}{+34.4}.
A 27B MLP run took 10.0--10.2 H100-hours (about 5.0\,h on two H100s), and a 4B run about 1.0--1.2\,h on one
H100 (\cref{tab:compute}). The socket needs one closed-form ridge fit per host. Computing the hosts' text
representations of the 9,481 genes takes 1--2.5\,min, and the fit with its 5-fold penalty search takes
5--9\,s. Scoring the primary arm (real and shuffled DNA) on all 2,972 development items takes
1.5--2\,min on one A100 for the hosts of 7B and smaller, and about 12\,min on one H100 for 27B
(\cref{tab:compute}).

\begin{figure}[h]
\centering
\includegraphics[width=0.85\linewidth]{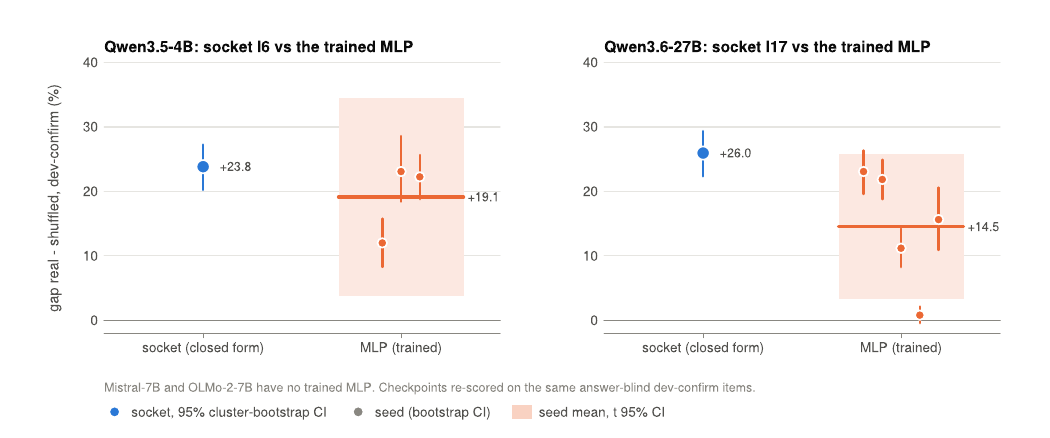}
\caption{\textbf{Socket versus the trained MLP adaptor} (\devconf{}). Dots: per-seed gaps of the MLP,
with 95\% bootstrap CIs. Bars: seed mean with $t$-interval. Line and band: the socket's gap and CI. Every
gap is real minus 5-let-shuffled DNA; the MLP passes both through its own trained connector.}
\label{fig:trained}
\end{figure}

\section{Modularity}
\label{app:modularity}

Test gaps for all twelve ordered pairs (source $\to$ host; own plug in brackets): into 4B
[$+22.2$]: 27B $+22.3$, \mistral{} $+22.4$, \olmo{} $+21.9$; into 27B [$+26.1$]: 4B $+26.4$,
\mistral{} $+26.1$, \olmo{} $+25.2$; into \mistral{} [$+16.7$]: 4B $+17.2$, 27B $+16.9$, \olmo{}
$+16.4$; into \olmo{} [$+18.4$]: 4B $+18.4$, 27B $+18.6$, \mistral{} $+18.5$.

\Devconf{} gaps with 95\% CIs: 27B plug into 4B $+22.0$ \ci{+18.8}{+25.1}; 4B plug into 27B $+26.5$
\ci{+22.7}{+30.1}; 4B and 27B plugs into \mistral{} $+17.4$ and $+16.9$; 4B, 27B and \mistral{} plugs
into \olmo{} $+19.9$ \ci{+16.3}{+23.2}, $+19.8$ \ci{+16.2}{+23.0} and $+20.2$ \ci{+16.5}{+23.4}.

\begin{figure}[h]
\centering
\includegraphics[width=0.8\linewidth]{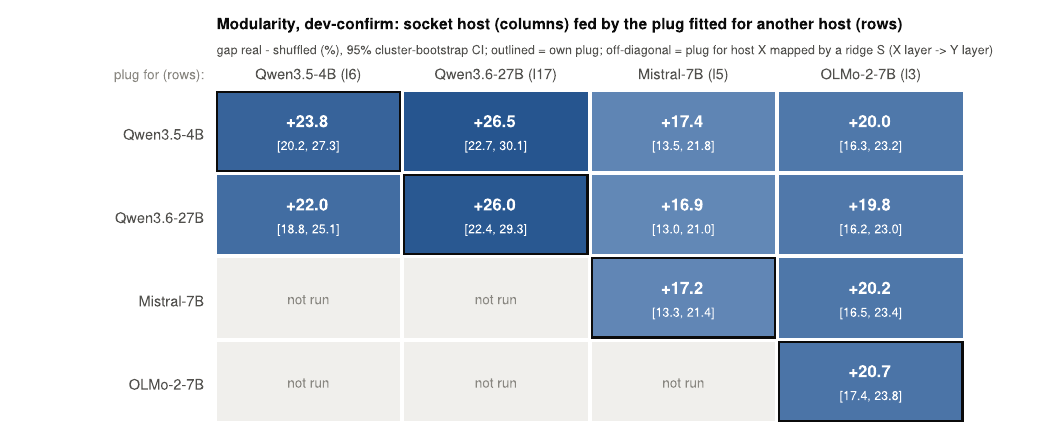}
\caption{\textbf{Modularity} (\devconf{}). Gap for each plug source (rows) and socket host (columns).
The diagonal is the host's own plug. Off-diagonal cells pass the source plug through a ridge map
between the hosts' text spaces. We did not run plugs from \mistral{} and \olmo{} into the Qwen hosts,
or from \olmo{} into \mistral{}.}
\label{fig:modularity}
\end{figure}

\section{What the socket carries: details}
\label{app:mechanism}

\paragraph{Homology, test.} With and without a training homolog (3,468 and 3,740 items), the plug
gives 4B $+23.9$ vs $+20.6$ (difference $+3.3$ \ci{-1.5}{+9.0}), 27B $+26.1$ vs $+26.1$ ($-0.1$
\ci{-3.7}{+3.4}), \mistral{} $+18.7$ vs $+14.9$ ($+3.8$ \ci{-1.1}{+9.4}) and \olmo{} $+19.7$ vs
$+17.2$ ($+2.5$ \ci{-2.3}{+7.4}). 1-NN gives $+23.1$ vs $+14.1$ ($+9.0$ \ci{+4.0}{+14.9}), $+27.9$ vs
$+16.6$ ($+11.4$ \ci{+6.0}{+17.4}), $+17.6$ vs $+10.1$ ($+7.5$ \ci{+2.3}{+14.0}) and $+16.9$ vs $+9.9$
($+7.1$ \ci{+2.4}{+12.3}). The refit and no-homolog lower bounds exceed 4 on every host, so the
wording rule does not apply. The description plug minus the description 1-NN is $-1.2$
\ci{-2.6}{+0.1}, $+3.1$ \ci{+1.5}{+4.8}, $+0.1$ \ci{-1.3}{+1.5} and $+2.1$ \ci{+0.6}{+3.4}. The plug refitted without
homologs is 4.7--6.6\% (point estimates) above the refitted 1-NN. On \devconf{} the 27B plug$-$1-NN
difference included zero, and the plug's homolog difference excluded zero on two hosts (below), so its
dependence on homologs is smaller than 1-NN's, not shown to be absent.

\begin{figure}[h]
\centering
\includegraphics[width=0.85\linewidth]{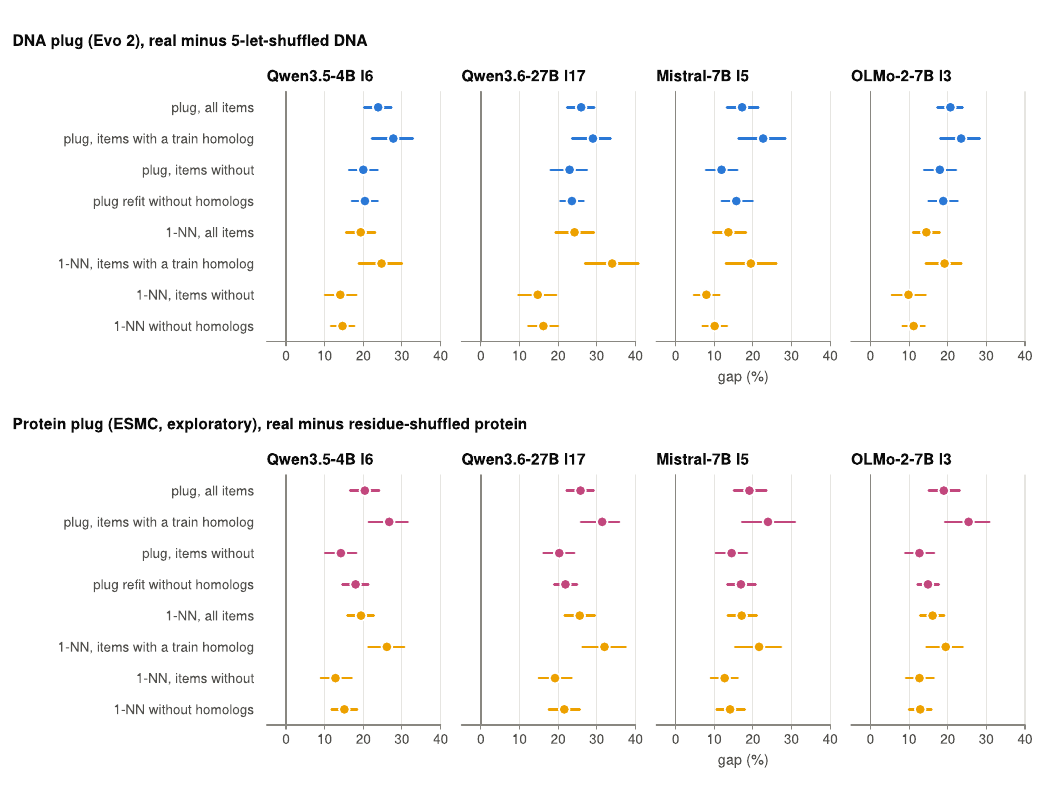}
\caption{\textbf{Homology} (\devconf{}). Plug and 1-NN gaps on all items, on items with and without a
training gene in the same homology group, and after refitting without any such training genes. Top:
the DNA plug (real minus 5-let-shuffled DNA). Bottom: the exploratory protein plug in the same frozen
sockets (real minus residue-shuffled protein; 1-NN in \esmc{} space; \cref{app:protein}).}
\label{fig:homology}
\end{figure}

\paragraph{Homology, \devconf{}.} Items whose gene has a training gene in the same homology group gain more than
items without (\cref{fig:homology}): 4B $+27.8$ vs $+20.0$ (difference $+7.8$ \ci{+0.9}{+14.0}), 27B
$+29.0$ vs $+23.0$ ($+6.0$ \ci{-1.0}{+12.2}), \mistral{} $+22.7$ vs $+11.9$ ($+10.8$ \ci{+3.5}{+17.5}),
\olmo{} $+23.5$ vs $+18.0$ ($+5.5$ \ci{-1.4}{+11.8}). Both the refit and the no-homolog lower bounds
exceed 4 on every host. So the plan's wording rule does not restrict the mechanism to ``transfer
from homologous genes''.

\paragraph{Lexical overlap.} On the 291 overlap items, the gain over sham is 4B $+34.0$ vs $+21.8$,
27B $+30.9$ vs $+23.7$ and \olmo{} $+21.6$ vs $+17.9$. The description-reading ceiling shows the same
split.

\paragraph{Test families and ceiling.} \fpath{} gains most ($+24.9$ to $+34.2$) and \fdis{} least ($+4.3$ to
$+9.0$). The plug reaches 86\%, 81\%, 84\% and 95\% of the mean-target ceiling (4B, 27B, \mistral{},
\olmo{}; \cref{tab:test-full}). On overlap items the gain over sham is $+19.3$ to $+31.8$, and
$+16.4$ to $+25.0$ on the rest.

\paragraph{\Devconf{} families and ceiling.} The gain is largest on \fpath{} ($+24.5$ to $+31.0$) and \fblock{} ($+13.6$ to
$+27.5$), and smallest on \fdis{} ($+6.9$ to $+15.9$). Writing the gene's own mean text representation
gives $+28.2$ (4B), $+33.6$ (27B), $+20.2$ (\mistral{}) and $+18.5$ (\olmo{}) over sham. The plug
reaches 86\%, 75\%, 92\% and 101\% of this ceiling (on \olmo{} the plug slightly exceeds it).

\paragraph{DNA as text.}
\label{app:dna-text}
Each of the 1,464 \devconf{} items is scored with its gene's MANE Select coding sequence (156 to
7,812\,bp on these genes) written as letters after ``\texttt{DNA:}'', where the slots would be. The
preamble reads ``Below is the coding sequence of a human gene, written as DNA letters. Read it, then
answer the question that follows.'' (plural for Shared-process items). The question, options and
answer instruction are those of the socket prompts, and no hook is set
(\texttt{scripts/dna\_as\_text.py}). The tokenizers read 1.8--1.9\,bp per token, and the longest
prompt has 5,443 tokens. Real, 5-let-shuffled (the socket's shuffles) and no-DNA accuracies are 4B
19.5, 19.3 and 19.7\%; 27B 22.1, 22.3 and 21.9\%; \mistral{} 20.8, 20.5 and 20.1\%; \olmo{} 20.8, 20.4
and 19.7\%. The gaps are $+0.2$ \ci{-0.8}{+1.3}, $-0.3$ \ci{-1.5}{+1.0}, $+0.3$ \ci{-0.4}{+1.2} and
$+0.3$ \ci{-0.4}{+1.1}, and real and shuffled sequences get the same letter on 88--94\% of items.
With ``\texttt{Gene: <symbol>}'' in place of the sequence, accuracy is 47.9\%, 58.7\%, 49.8\% and
40.9\%, which shows that the scoring reads the answers. For comparison, the socket's real-DNA
accuracy on the same items is 43.6\%, 47.8\%, 38.7\% and 40.3\%.

\section{Two larger hosts: mutant and non-human genes}
\label{app:larger}

These checks are exploratory. They ran once, after the test run, outside the frozen plan, on two hosts that
are not among the four: \qbig{} (64 blocks, width 5,120) and \dsflash{} (DeepSeek-V4-Flash-0731, 284B
parameters, mixture of experts, 43 blocks). \dsflash{} answers more questions from the gene name alone:
63.7\% against 56.4\% on the 1,464 \devconf{} items (question alone 20.2\% and 21.2\%). \dsflash{} keeps
four residual streams of width 4,096 per token; its state after a block is the four joined (16,384
numbers), which the text targets, the plug and the replace write all use, each stream norm-matched on
its own. Its FP8/FP4 weights were converted to bf16, and its blocks spread over eight 80\,GB A100s. For
each host the half-max rule chose the layer (block 15 and block 7), and the text targets and plug were
made as for the four hosts. Nothing was refitted for the checks below.

\paragraph{Mutant DNA} (\devconf{}, \texttt{scripts/mutant\_dna.py}). Each \devconf{} gene's coding
sequence is rewritten four ways, read by \evo{} and written through the same plug: \emph{synonymous}
(every inner codon replaced by the synonymous codon sharing the fewest letters with it; 57.5\% of letters
kept, same protein), \emph{frameshift} (letter 31 deleted; the protein is nonsense after residue 10),
\emph{early stop} (codon 11 replaced by TAA) and \emph{reversed} (read backwards). \Cref{tab:mutant} gives
the results. Frameshift and early stop do not differ from real DNA (paired, $+0.3$ and $-0.2$ on \qbig{},
$+0.2$ and $-0.3$ on \dsflash{}, every CI within $\pm1.2$). Synonymous recoding costs $-4.6$ \ci{-6.5}{-2.8}
and $-2.8$ \ci{-4.7}{-0.8}. Reversed DNA is at the shuffled level.

\begin{table}[h]
\centering
\small
\setlength{\tabcolsep}{3.5pt}
\caption{\textbf{Rewritten DNA} (1,464 development questions, two larger models). Accuracy (\%) when each gene's DNA is
rewritten before it is written into the model, and the gap to shuffled DNA (\%, with 95\% confidence interval).
Frameshift and early stop destroy the protein but barely change the DNA; synonymous recoding keeps the protein
but changes 42\% of the letters; reversed reads the DNA backwards.}
\label{tab:mutant}
\begin{tabular}{@{}lrlrl@{}}
\toprule
 & \multicolumn{2}{c}{\qbig{}} & \multicolumn{2}{c}{\dsflash{}} \\
DNA written & acc. & gap & acc. & gap \\
\midrule
Real & 44.1 & $+23.4$ {\scriptsize\ci{20.0}{26.8}} & 48.2 & $+28.2$ {\scriptsize\ci{24.1}{32.2}} \\
Frameshift & 44.4 & $+23.6$ {\scriptsize\ci{20.2}{27.0}} & 48.4 & $+28.4$ {\scriptsize\ci{24.3}{32.5}} \\
Early stop & 43.9 & $+23.2$ {\scriptsize\ci{19.5}{26.6}} & 48.0 & $+27.9$ {\scriptsize\ci{23.8}{31.9}} \\
Synonymous & 39.5 & $+18.7$ {\scriptsize\ci{15.2}{22.2}} & 45.4 & $+25.4$ {\scriptsize\ci{21.4}{29.4}} \\
Reversed & 19.5 & $-1.2$ {\scriptsize\ci{-3.4}{0.9}} & 19.9 & $-0.1$ {\scriptsize\ci{-2.7}{2.6}} \\
Shuffled & 20.8 & & 20.0 & \\
Zero write & 21.4 & $+0.7$ {\scriptsize\ci{-2.1}{3.4}} & 20.6 & $+0.5$ {\scriptsize\ci{-2.0}{3.1}} \\
\bottomrule
\end{tabular}
\end{table}

\paragraph{Non-human genes} (\texttt{scripts/exotic\_genes.py}, \texttt{data/exotic/}). The genes are
reviewed UniProt proteins of \textit{Saccharomyces cerevisiae}, \textit{Arabidopsis thaliana},
\textit{Escherichia coli} K-12 and the archaeon \textit{Methanocaldococcus jannaschii}, each with a coding
sequence that translates to the protein. Each gene is aligned against all 18,758 human proteins of the
study with the exact aligner of the split and sorted into a group: \emph{visible} (a human protein aligns
with $E\le10^{-3}$), \emph{remote} (no alignment with $E\le0.1$, but a shared InterPro homologous
superfamily, domain or family) or \emph{none} (no alignment with $E\le0.1$ and no shared InterPro entry).
InterPro entries are themselves profiles, so \emph{remote} means invisible to pairwise alignment but
visible to profile methods. The questions follow \geneblind{}'s answer-blind rule, one per gene and
family, at most 300 per family and group: the gene's GO-slim biological process (negatives as in
\fblock{}) or, for enzymes, its reaction subclass (the first two EC numbers, with published experimental
evidence). There are 1,635 questions about 1,557 genes (600 visible, 600 remote, 435 none); the archaeon
contributes only 57 enzyme questions. The arms are the gene's DNA, its 5-let shuffle, the DNA of another
gene of the same organism, group and family, 1-NN (the text target of the cosine-nearest human fit gene
in \evo{} space, as in \cref{sec:res-mechanism}), and three prompts without DNA: the gene's name with its
organism, the organism alone and the question alone. As a sequence-search reference we copy the
annotation of the best-aligned human protein (a uniform pick among the options it has, else among all five).

\begin{table}[t]
\centering
\footnotesize
\setlength{\tabcolsep}{3.5pt}
\caption{\textbf{Non-human genes}, grouped by how related they are to the human genes the map was fitted on:
\emph{visible}, a human relative found by alignment; \emph{remote}, only a shared protein domain; \emph{none}, no
relation found. Accuracy (\%; chance 20\%). Real / Shuf.: the gene's DNA or a shuffled copy. Other: the DNA of
another gene of the same organism and group. 1-NN: the nearest human training gene written instead; 1-NN shuf.: its
shuffled control; Plug $-$ 1-NN: the plug's lead over it (\%, 95\% confidence interval). Name: the gene's name
and organism given as text, no DNA. Organism: the organism's name alone. Search: copying the annotation of the
best-aligned human protein. An empty write and the question alone give 17--24\%.}
\label{tab:exotic}
\begin{tabular}{@{}llrrrrrlrrr@{}}
\toprule
Host & Group ($n$) & Real & Shuf. & Other & 1-NN & 1-NN shuf. & Plug $-$ 1-NN & Name & Organism & Search \\
\midrule
\multirow{4}{*}{\qbig{}} & visible (600) & 34.8 & 17.3 & 22.0 & 39.2 & 16.3 & $-4.3$ {\scriptsize\ci{-8.2}{-0.5}} & 45.8 & 26.0 & 69.6 \\
 & remote (600) & 29.8 & 17.7 & 20.3 & 28.5 & 17.7 & $+1.3$ {\scriptsize\ci{-2.7}{5.2}} & 41.2 & 20.3 & 23.7 \\
 & none (435) & 35.6 & 18.9 & 22.1 & 23.2 & 16.8 & $+12.4$ {\scriptsize\ci{7.3}{17.7}} & 45.1 & 25.1 & 20.3 \\
 & all (1,635) & 33.2 & 17.9 & 21.4 & 31.0 & 16.9 & $+2.2$ {\scriptsize\ci{-0.3}{4.7}} & 43.9 & 23.7 & 39.7 \\
\midrule
\multirow{4}{*}{\dsflash{}} & visible (600) & 47.8 & 19.0 & 19.8 & 51.8 & 16.3 & $-4.0$ {\scriptsize\ci{-8.3}{0.3}} & 68.3 & 25.2 & 69.6 \\
 & remote (600) & 36.7 & 20.2 & 18.5 & 32.0 & 19.7 & $+4.7$ {\scriptsize\ci{0.0}{9.2}} & 65.8 & 24.0 & 23.7 \\
 & none (435) & 39.5 & 18.9 & 23.9 & 26.4 & 17.5 & $+13.1$ {\scriptsize\ci{7.7}{18.9}} & 65.8 & 27.8 & 20.3 \\
 & all (1,635) & 41.5 & 19.4 & 20.4 & 37.8 & 17.9 & $+3.7$ {\scriptsize\ci{0.8}{6.6}} & 66.7 & 25.4 & 39.7 \\
\bottomrule
\end{tabular}
\end{table}

\Cref{tab:exotic} gives the results. Over all questions the plug beats shuffled DNA by $+15.3$
\ci{12.9}{17.7} (\qbig{}) and $+22.1$ \ci{19.1}{25.1} (\dsflash{}), while the DNA of another gene of the
same organism, group and family gives 18.5--23.9\%, near the shuffled level (17.3--20.2\%), so the gain is
specific to the gene, not to the organism. Where a human relative aligns, copying its annotation (69.6\%) is far ahead of the socket,
and 1-NN is ahead of the plug. Where none exists, sequence search and 1-NN fall to near chance, while the
plug keeps 36--40\%.

\paragraph{Distance from the human genes.} In the space the plug reads (pooled \evo{} features,
standardised with the plug's standardiser), the median cosine of a non-human gene to its nearest human
gene is 0.38 (yeast 0.46, \textit{E.~coli} 0.32), against 0.65 for a human gene to its nearest human gene
of another homology group; 69\% of the non-human genes lie below the human genes' 10th percentile. The
nearest human gene in this space is the best-aligned human protein for only 41\% of the visible genes.
Plug accuracy falls with this distance but stays above the other-gene control in the farthest third
(\dsflash{} 44.6, 42.8 and 37.2\% by thirds, \qbig{} 35.8, 34.5 and 29.4\%). The plug moves these genes
only partly toward the human genes: among the 15 nearest neighbours of a non-human gene (cosine, all 11,038
genes, 86\% of them human), 15\% are human genes before the plug and 33\% (\qbig{}) and 24\% (\dsflash{}) in
the vectors the plug writes; for genes with no human relative 11\%, 24\% and 15\%. Most neighbours stay
non-human genes of the same organism (79\% before, 60\% and 69\% after).

\section{Rotation against ridge}
\label{app:rotation}

The rotation is the map of \citet{Cheng2026-PlatosKey}: on centred paired rows, $R=UV^\top$ from the thin SVD
$X^\top Y=U\Sigma V^\top$ (orthogonal Procrustes, \citealp{Schonemann1966-Procrustes}; semi-orthogonal when the
widths differ), and one global scale that gives the mapped rows the total spread of the targets. It is fitted
on the plug's own 8,319 training genes and scored in the same socket at the same layer, on the 1,464
\devconf{} items, with and without the plug's norm match on top. \emph{DNA $\to$ host}: the rotation from the
standardised \evo{} features to the host's mean text target, in place of the ridge plug. \emph{Host $\to$
host}: the other host's ridge plug, carried into this host by a map between the two hosts' text targets,
fitted on text only, ridge (as in \cref{sec:res-modularity}) or the rotation; the other host is \qfour{}, and
\qtwentyseven{} for \qfour{} itself. The two 27B hosts ran on two 40\,GB A100s; \qtwentyseven{} is bit for bit
only on an H100, so its ridge gap here is $+25.4$ against the recorded $+26.0$ (\cref{app:repro}). Every
ridge and rotation arm's shuffled accuracy and the zero write lie at 19--23\%.

\begin{table}[h]
\centering
\footnotesize
\setlength{\tabcolsep}{3.2pt}
\caption{\textbf{Rotation against ridge} (1,464 development questions). Gap (\%) for two ways of translating
between models. DNA $\to$ host: from the DNA model into the language model, by our ridge map or by a rotation.
Host $\to$ host: another language model's plug carried into this one by a map fitted on text alone, ridge or
rotation. Rot.\ + NM: the rotation with each written vector set to the typical length (our norm match).
Differences to ridge with confidence intervals are in the text.}
\label{tab:rotation}
\begin{tabular}{@{}lrrrrrr@{}}
\toprule
 & \multicolumn{3}{c}{DNA $\to$ host} & \multicolumn{3}{c}{host $\to$ host} \\
\cmidrule(lr){2-4}\cmidrule(l){5-7}
Host & ridge & rotation & rot.\ + NM & ridge & rotation & rot.\ + NM \\
\midrule
\qfour{} & +23.8 & +9.4 & +11.5 & +22.0 & +15.0 & +22.1 \\
\qtwentyseven{} & +25.4 & +16.5 & +16.5 & +26.3 & +24.4 & +27.3 \\
\mistral{} & +17.2 & +10.0 & +10.2 & +17.4 & +14.1 & +17.2 \\
\olmo{} & +20.7 & +10.0 & +11.8 & +20.0 & +20.3 & +20.8 \\
\qbig{} & +23.4 & +12.6 & +13.1 & +23.5 & +19.2 & +23.1 \\
\bottomrule
\end{tabular}
\end{table}

Rotation minus ridge, DNA $\to$ host: $-14.3$ \ci{-17.2}{-11.6} (\qfour{}), $-8.7$ \ci{-11.1}{-6.4} (\qtwentyseven{}),
$-8.0$ \ci{-10.2}{-5.9} (\mistral{}), $-8.8$ \ci{-11.5}{-6.2} (\olmo{}) and $-10.7$ \ci{-13.1}{-8.3} (\qbig{}); with
the norm match $-11.7$, $-8.3$, $-6.8$, $-7.6$ and $-10.2$, every CI below zero. Host $\to$ host, rotation with
norm match minus ridge: $+0.1$ \ci{-1.0}{+1.2}, $+0.5$ \ci{-0.6}{+1.5}, $-0.3$ \ci{-1.2}{+0.5}, $+0.5$ \ci{-0.8}{+2.2}
and $-0.3$ \ci{-1.4}{+0.8}; without it $-8.1$, $-1.8$, $-2.8$, $-0.1$ and $-4.6$. On its training genes the ridge plug
reproduces 39--59\% of the target variance ($R^2$), the rotation none ($R^2$ $-0.11$ to $-0.17$). \dsflash{} has not
been run.

\section{Second modality: protein details}
\label{app:protein}

\begin{table}[h]
\centering
\small
\setlength{\tabcolsep}{3.2pt}
\caption{\textbf{Protein instead of DNA} (development data; not part of the pre-registered test). The same frozen
models and layers as \cref{tab:test}, with a map from the protein model \esmc{} in place of the DNA map. Gap (\%):
real minus shuffled protein, with 95\% confidence interval; every model passes the test corrected for four models.
1-NN: the nearest training gene in \esmc{}'s space written instead. DNA gap: the DNA map in the same run.
Protein $-$ DNA and Both $-$ DNA: accuracy differences to the DNA map alone.}
\label{tab:protein}
\begin{tabular}{@{}lcccc@{}}
\toprule
 & 4B & 27B & Mistral & OLMo \\
\midrule
\textbf{Protein gap} & \textbf{+20.4} & \textbf{+25.8} & \textbf{+19.1} & \textbf{+19.0} \\
\quad 95\% CI & {\scriptsize\ci{16.6}{24.1}} & {\scriptsize\ci{22.2}{29.2}} & {\scriptsize\ci{15.1}{23.5}} & {\scriptsize\ci{15.1}{23.1}} \\
Gap, nearest training gene & +19.4 & +25.6 & +17.1 & +16.1 \\
Plug $-$ nearest training gene & +1.0 & +0.2 & +2.1 & +2.9 \\
\quad 95\% CI & {\scriptsize\ci{-2.4}{4.0}} & {\scriptsize\ci{-3.5}{3.8}} & {\scriptsize\ci{-0.8}{5.0}} & {\scriptsize\ci{-0.1}{5.8}} \\
Gap, refitted without relatives & +18.0 & +21.9 & +16.9 & +14.9 \\
DNA gap & +23.8 & +25.8 & +17.2 & +20.7 \\
Protein $-$ DNA, real accuracy & $-0.9$ & +2.1 & +0.9 & $-0.1$ \\
Both $-$ DNA, real accuracy & +1.1 & +1.0 & +1.0 & +0.2 \\
\bottomrule
\end{tabular}
\end{table}

\paragraph{Summary.} We plugged a second encoder into the same frozen sockets and layers. The plug is a ridge map from
\esmc{} features of each gene's protein, fitted exactly like the DNA plug. The null is a residue
shuffle of the protein. This analysis is exploratory: we developed it on \devtune{} and confirmed it on
\devconf{}; it is not pre-registered and never touched the test set. On \devconf{}, the protein plug gives
$+19.0$ to $+25.8$\% (\cref{tab:protein}). The permutation, gene-swap and random-map controls
are null (each permuted plug's own gap lies between $-2.3$ and $+3.3$). Through the same text-only
maps, the protein plug crosses hosts within 1.4\% of the host's own plug. Protein and DNA carry
about the same usable signal. Writing both gains at most about 1\% over DNA alone.

The main caveat is that \emph{the protein plug is about as good as 1-NN retrieval in \esmc{}
space}. The paired difference is $+0.2$ to $+2.9$\%, and every CI includes zero. Unlike the DNA
plug, the protein plug shows no advantage over retrieval on most hosts. A plug refitted without
homologs keeps $+14.9$ to $+21.9$\%. It beats a 1-NN without homologs by $+2.1$ to $+4.1$, with
CIs that exclude zero on three hosts (below). We read this as evidence that the socket
accepts a second modality at about the level of nearest-homolog retrieval. We do not read it as a
learned mapping from protein to text.

\paragraph{Setup.} The protein of a gene is the translation of its MANE Select coding sequence. Its
feature is the \esmc{} 600M layer-30 representation: means over 16-residue bins, then mean-pooled in
float64 exactly as the socket pools \evo{}, and z-scored with training statistics. 9,481 genes have a
protein. The encoder sees 2,046 residues, so we truncate the 147 proteins that are longer. The null
shuffles the residues the encoder sees, seeded per gene, so the composition of the encoder input
stays exactly the same. The plug, norm match, sockets, layers, controls and bootstrap are those of
the DNA analysis, and the plug is fitted on the same 8,319 training genes. We developed the protein
plugs on \devtune{} and refitted them for \devconf{} with identical weights (hash-checked). Holm uses $m=4$
hosts at one-sided $\alpha=0.025$. \Devtune{} gaps were $+20.8$, $+25.1$, $+16.6$ and $+16.1$. The DNA arm
in this run uses the same pack as \cref{tab:main}, but on 27B it scores the \devconf{} items without the
\devtune{} items in the same batches. So its 4B gap ($+23.8$) equals \cref{tab:main}, and its 27B gap
($+25.8$) is 0.2\% lower (27B batch composition; \cref{app:deviations}).

\begin{table}[h]
\centering
\small
\caption{\textbf{Protein instead of DNA, all rows} (development data). Gaps (\%) are real minus shuffled protein
unless the row says otherwise; controls are given against the empty write (sham).}
\label{tab:protein-full}
\begin{tabular}{@{}lcccc@{}}
\toprule
 & \qfour{} $\ell6$ & \qtwentyseven{} $\ell17$ & \mistral{} $\ell5$ & \olmo{} $\ell3$ \\
\midrule
Accuracy: real / shuffled (\%)        & 42.7 / 22.3 & 49.7 / 24.0 & 39.5 / 20.4 & 40.2 / 21.2 \\
Gap [95\% CI]                     & +20.4 \ci{16.6}{24.1} & +25.8 \ci{22.2}{29.2} & +19.1 \ci{15.1}{23.5} & +19.0 \ci{15.1}{23.1} \\
Real $-$ empty write                     & +23.4 & +27.0 & +19.5 & +18.5 \\
Real $-$ another gene's protein                & +22.9 & +29.2 & +18.6 & +21.3 \\
Real $-$ re-paired plugs (mean of 5)     & +22.8 & +29.1 & +18.7 & +20.4 \\
Real $-$ random map               & +24.4 & +28.8 & +20.6 & +21.2 \\
Gap of each re-paired plug      & $-1.8$\ldots$+3.3$ & $-2.3$\ldots$+2.3$ & $-2.3$\ldots$+1.6$ & $-1.8$\ldots$+3.2$ \\
Plug $-$ nearest gene, real accuracy           & +1.1 \ci{-1.4}{3.4} & +0.9 \ci{-1.5}{3.2} & +0.7 \ci{-1.3}{2.6} & +3.8 \ci{1.6}{5.8} \\
Plug $-$ nearest gene, gap    & +1.0 \ci{-2.4}{4.0} & +0.2 \ci{-3.5}{3.8} & +2.1 \ci{-0.8}{5.0} & +2.9 \ci{-0.1}{5.8} \\
Plug, genes with / without relatives & +26.7 / +14.2 & +31.4 / +20.3 & +23.9 / +14.5 & +25.4 / +12.7 \\
Nearest gene, with / without relatives      & +26.1 / +12.8 & +32.0 / +19.2 & +21.6 / +12.7 & +19.5 / +12.7 \\
DNA plug, with / without relatives  & +27.8 / +20.0 & +28.9 / +22.7 & +22.7 / +11.9 & +23.5 / +18.0 \\
Plug refitted without relatives           & +18.0 \ci{14.6}{21.3} & +21.9 \ci{19.0}{24.9} & +16.9 \ci{13.4}{20.7} & +14.9 \ci{12.2}{17.7} \\
Nearest gene without relatives                 & +15.1 & +21.6 & +14.1 & +12.9 \\
Refit $-$ nearest gene, both without relatives & +4.1 \ci{1.3}{6.8} & +2.1 \ci{-1.0}{5.2} & +2.5 \ci{0.1}{4.9} & +2.9 \ci{0.2}{5.8} \\
Protein $-$ DNA, real accuracy            & $-0.9$ \ci{-2.9}{1.1} & +2.1 \ci{0.0}{4.0} & +0.9 \ci{-0.9}{2.6} & $-0.1$ \ci{-2.0}{1.7} \\
Gap, both encoders             & +23.4 & +24.8 & +18.5 & +18.3 \\
Both $-$ DNA, real accuracy               & +1.1 \ci{-0.3}{2.5} & +1.0 \ci{-0.5}{2.4} & +1.0 \ci{-0.4}{2.5} & +0.2 \ci{-1.1}{1.5} \\
Both $-$ protein, real accuracy           & +2.0 \ci{0.4}{3.7} & $-1.1$ \ci{-2.5}{0.2} & +0.1 \ci{-1.1}{1.4} & +0.3 \ci{-1.0}{1.7} \\
\bottomrule
\end{tabular}
\end{table}

\paragraph{Modularity.} Gaps of the protein plug passed through another host's text space (real
minus shuffled): into 4B via 27B $+21.1$; into 27B via 4B $+24.6$; into \mistral{} via 4B and 27B
$+18.3$ and $+19.4$; into \olmo{} via 4B, 27B and \mistral{} $+20.4$, $+19.9$ and $+19.5$. Each lies
within 1.4\% of the host's own protein plug.

\paragraph{Compute.} Scoring all protein arms on \devconf{} took 507\,s (4B, plus 102\,s for its modularity
run), 738\,s (\mistral{}) and 846\,s (\olmo{}) on one A100 each, and 870\,s plus 242\,s (27B) on one
H100.

\section{Statistics and the frozen plan}
\label{app:stats}

\paragraph{Endpoint.} The gap is accuracy with real DNA minus accuracy with shuffled DNA, pooled over
all four families (item-weighted), in \% of the items, at each host's test layer with the primary plug. The gap without \fdis{} is a
sensitivity analysis, because the rule that decides whether to keep \fdis{} has not run.

\paragraph{Inference.} Each gap gets a 95\% percentile interval from a cluster bootstrap (2,000
resamples, seed 20261012). The cluster of an item is the homology group of its first gene. The test is
one-sided ($H_1$: gap $>0$; bootstrap $p$ = share of resamples $\le0$), with Holm correction over the
hosts at one-sided $\alpha=0.025$. A host is ``above SESOI'' when the lower bound exceeds 4\%.
Clusters are not fully independent: an \fshare{} item's second gene can belong to another cluster. Resampling
genes instead of first-gene clusters gave narrower intervals in a check, so the first-gene cluster
interval is the more conservative choice. The homology-split difference resamples the two strata
independently, as the plan's stratified bootstrap prescribes. Secondary CIs (plug minus 1-NN, homology
differences) are not corrected for multiple hosts.
We summarise the trained MLP per seed and by a $t$ interval over seed means. The ridge penalty grid is
nine multiples, $10^{-4}$ to $10^{2}$, of the mean eigenvalue of the centred Gram matrix.

\paragraph{The frozen plan.} The test run is a single pass over all four hosts. It uses the plugs
from the development fits (file hashes checked, never refitted), Holm over four hosts, and every arm above.
It is the only evaluation that is clean for every design choice. \Devconf{} numbers for 4B and 27B confirm
the item build and the layer; they are not clean for the plug design. One wording in the frozen plan is
wrong and is corrected here, not in the plan: its section 7 calls the trained arms' floor a trained
NULL-DNA arm, but the analysis it specifies, and the one we ran, measures every trained gap as real
minus 5-let-shuffled DNA through the same connector (\cref{app:trained}).

\section{Deviations and disclosures}
\label{app:deviations}

\begin{itemize}
  \item \textbf{Borrowed shuffles (\devconf{}, 4B and 27B).} In the confirmatory run, 2,147 of the 4,158
  development DNA slots used a 5-let shuffle of the same gene taken from the earlier development
  pack. A later pack (\texttt{f857cdc2}) gives every slot its own shuffle. We regenerated the 2,007
  shuffles shared with the earlier pack, and they agree exactly. \Devtune{} values agree within 0.4\%
  between packs. All 4B and 27B \devconf{} numbers in this paper use \texttt{f857cdc2}, including the
  trained MLP, which we re-scored on it; the released code reproduces them. On the earlier pack, the
  confirmatory run gave $+23.5$ \ci{+20.0}{+26.8} (4B) and $+26.0$ \ci{+22.4}{+29.4} (27B), with the
  same Holm decisions, and the MLP seed means were $+18.7$ (4B) and $+14.7$ (27B). No decision was made on \texttt{f857cdc2}: it was adopted after
  the test, so that the paper and the released code report the same numbers. \mistral{} and \olmo{}
  used their own shuffles throughout. Every test slot gets its own shuffle.
  \item \textbf{Cluster fallback.} Splitting by connected components put 83\% of development items
  into one component, above the 25\% cap. We therefore assigned items by their first gene's homology
  group, and dropped 381 items whose genes fell in both halves (split manifest sha256
  \texttt{76a2865e\ldots}). The test bootstrap uses the same first-gene clusters.
  \item \textbf{Training schedule of the trained MLP.} The trainer was meant to run the description stage once
  and then the question stage for the remaining 4,557 steps. Its sampler instead starts a new epoch, the description stage
  first, after 4,477 steps. All eight re-scored checkpoints therefore ran the description stage (2,838 steps), the question stage (1,639), the description stage again (2,838) and
  80 question-stage steps, as their loss curves show (a jump at step 4,477). The question stage got 1,639 steps at
  learning-rate multipliers of 0.73 to 0.37 of the peak, then 80 steps near zero (1,719 in all). We did not retrain. The MLP may therefore be below what they would reach with the intended schedule.
  \item \textbf{Half-max rule in sample.} The half-max rule was chosen from 36 candidates by regret on the three
  development hosts. Its agreement with these hosts is not evidence for the rule.
  \item \textbf{\olmo{} sanity oracle.} Before we committed \olmo{}'s prediction, a sanity check
  wrote its description representation at $\ell=4$, 12 and 20 ($+14.4$, $+0.8$, $+0.1$). The
  prediction follows mechanically from the committed probe file under the frozen rule, and does not
  use these numbers.
  \item \textbf{\olmo{} inclusion after its development results.} \olmo{} was chosen as the fourth
  host by a fixed rule (no gated weights, a new model family; fallback Phi-3.5-mini-instruct), and its
  half-max-rule prediction was committed before any plug scoring. The \olmo{} amendment, which adds \olmo{} to the
  test plan, was approved after its \devtune{} curve and \devconf{} result ($+20.7$) were known. Had \olmo{}
  failed on development data, it might have been replaced. The layer prediction is still out of
  sample, and the \olmo{} test result comes from test items that no one had seen. But the choice to
  include this host was not blind to its development results.
  \item \textbf{Pipeline reproduction.} Before the test, the consolidated pipeline reproduced every
  recorded \devconf{} number (gaps, CIs, controls, splits, maps) to within $10^{-6}$. This needed three
  fixes: a virtual-environment wrapper, the 27B text-representation batch size and a pack-hash
  check. We re-ran \olmo{} on the fixed pipeline, and it matched exactly.
  \item \textbf{Test build path.} The first test-item build crashed \emph{after} sealing the items,
  while copying the manifest record into the read-only deployed code tree. No item was read and no
  score existed. A fix that writes the record next to the test data changed only the build path (one
  build script and its test; no scoring or analysis module imports it); the partial build was
  deleted unread and rebuilt with the identical command and seed. The builder is deterministic, but
  the crashed build's manifest hash was never captured, so identity with the rebuild is not verified.
  Two further launches of the \evo{} feature extraction for the test slots failed before any feature
  was written: a missing environment variable, then an output directory inside the read-only tree.
  A wrapper script that sets both fixed them, and the third launch ran. The test run used the fixed
  code on read-only deployed trees at both sites, under the plan with the two amendments. No arm was added,
  dropped or tuned after the test items were built.
  \item \textbf{Post-run reads.} The test gate refuses any read of test items outside the run, but
  after the run it did not block a new scoring read that names the run's manifest. Such reads are
  auditable rather than blocked. An audit of both compute clusters found no result file written after the run
  started outside the run's own directories, and each run directory holds exactly the planned score
  files, so no such read occurred.
  \item \textbf{27B batch composition.} On 27B, which items share a batch changes the result of
  1--8 of 1,464 \devconf{} items per arm (\devconf{} items scored alone: $+25.8$; scored together with \devtune{}
  items: $+26.0$). The other hosts flip no item. The test items were scored alone, as in the rehearsal.
  \item \textbf{Ceiling rows.} The description-reading ceilings of test genes use text representations
  computed in the run (eval slice). These differ from the frozen slice rows at bf16 level, which
  flips 1--9 of 1,464 ceiling items in the rehearsal. Only the ceilings are affected.
  \item \textbf{Modular map space.} The plan describes the text-to-text map as fitted from host $X$'s
  description space; the code fits it on the mean target (symbol, name and description), as
  \cref{sec:meth-socket} states.
  \item \textbf{Discarded 4B \devconf{} pass.} A first 4B \devconf{} pass used a pack built on the other cluster,
  whose \evo{} features differ numerically (cosine 0.99999). It was discarded and rerun on the
  reference pack (both on the earlier pack: $+23.8$, then $+23.5$); nothing was chosen on it.
  \item \textbf{Re-generated analyses.} The per-host test analyses of the three smaller hosts were
  regenerated by the Holm job from the same score files; they equal the originals by determinism
  (fixed bootstrap seed), not by a recorded comparison.
  \item \textbf{Test transfer.} The test pack went from one compute cluster to the other through a
  stream that wrote nothing to the workstation's disk; its SHA-256 sum matched at both sites.
  \item \textbf{Protein arm.} Not pre-registered; developed on \devtune{} and confirmed on \devconf{}; no test
  item was built or read for it.
  \item \textbf{DNA as text.} Added after the test run, on \devconf{} only (\cref{app:dna-text}). 27B
  ran on two 40\,GB A100s with the A100 kernel settings, so its numbers are close to, not bit-equal
  with, an H100 run.
\end{itemize}

\paragraph{Exploratory history.} On the older development set, the first socket (description plug)
gave $+13.0$ (4B $\ell=6$) and $+21.4$ (27B $\ell=17$). The improved plug (mean target, norm match)
gave $+22.6$ and $+27.5$. These runs used different option sets and clusters, so they are not
replications.

\section{Hardware and reproducibility}
\label{app:repro}

Socket fits and scoring ran on NVIDIA A100 (40\,GB) and H100 GPUs. The host and the encoder are
never trained. \Cref{tab:compute} lists the recorded wall-clock time, GPU utilisation and memory per
stage. We record hashes for all plugs, slices and item manifests. The frozen plugs for every test and
descriptive layer are stored with identical SHA-256 sums on both compute clusters. Ridge fits are
float64 and deterministic. Scoring repeated in separate jobs with the same batch plan is
bit-identical. On 27B, a different batch composition flips up to 0.5\% of items per arm
(\cref{app:deviations}). The 27B numbers are bit-identical only on an H100. Re-run once on an A100
(80\,GB) with the A100 kernel settings, the 27B \devconf{} gap was $+25.4$ \ci{+21.9}{+28.8} against
the recorded $+26.0$, with the same Holm and SESOI decisions, and 96.1--98.7\% of answer letters were
equal per arm. The other three hosts are bit-identical on the A100.

\begin{table}[h]
\centering
\footnotesize
\setlength{\tabcolsep}{2.1pt}
\caption{\textbf{Compute summary.} Wall-clock time per job or step, as recorded in the run logs.
GPU utilisation is the mean of \texttt{nvidia-smi} samples while the job held the GPU (from the
first busy sample where stated). Reps: the host's pooled symbol, name and description
representations of the 9,481 genes at the socket layer. Fit: ridge with the 5-fold, nine-penalty
search on 8,319 genes. Primary: real and shuffled DNA on all 2,972 development items. The 4B and
27B fit and primary-arm times come from the reproduction of the \devconf{} run on the consolidated
pipeline, which matched it exactly (\cref{app:deviations}). n/r: not recorded.}
\label{tab:compute}
\begin{tabular}{@{}llllll@{}}
\toprule
Stage & Host & GPU & Wall-clock & GPU util. & Memory \\
\midrule
Reps / fit / primary & \qfour{} $\ell6$ & 1 A100 & 67\,s / 8\,s / 105\,s & 61\% (job) & 22\,GB (job) \\
 & \qtwentyseven{} $\ell17$ & 1 H100 & 136\,s / 5\,s / 700\,s & 55--79\% scoring$^c$ & n/r \\
 & \mistral{} $\ell5$ & 1 A100 & 78\,s / 9\,s / 91\,s & 44\% (reps job) & 17\,GB (reps job) \\
 & \olmo{} $\ell3$ & 1 A100 & 81\,s / 9\,s / 99\,s & n/r & n/r \\
\midrule
\Devconf{} & \qfour{} & 1 A100 & 260--365\,s $\times$ 11 tasks & 61--70\% & 16--22\,GB \\
 & \qtwentyseven{} & 1 H100 & 620--856\,s $\times$ 5 tasks & 62--75\% & 66\,GB \\
 & \mistral{} & 1 A100 & 127--910\,s $\times$ 4 tasks$^a$ & 53--80\% & 25--28\,GB \\
 & \olmo{} & 1 A100 & 132--900\,s $\times$ 5 tasks & ${\approx}100\%$ & 29\,GB \\
\midrule
\Devtune{} curve & \mistral{} $\ell$0--30 & 1 A100 & 30 + 28\,min & 81\%, 78\% & 26\,GiB \\
 & \olmo{} $\ell0$--30 & 2 A100 & 32\,min & 100\% (23/32 samp.) & 27\,GB \\
\midrule
Protein, \devconf{} & all four & A100; H100 (27B) & 507--1,112\,s per host & n/r & n/r \\
\midrule
Trained (training) & 4B MLP & 1 H100 & 1.16\,h (s3); 4.88\,h (s1--2)$^b$ & n/r & 21\,GiB \\
 & 27B MLP & 2 H100 & 4.9--5.0\,h (10.0--10.2\,GPU-h) & n/r & 37\,GiB/GPU \\
Trained (re-score) & 4B, 27B & A100 / H100 & ${\approx}120$\,s / 270--380\,s & 56--76\% / 36--50\% & n/r \\
\midrule
Test, all arms & 4B, Mistral, OLMo & 4 A100 & 69\,min, one job & n/r & n/r \\
 & 27B & 4 H100 & 64\,min, one step & n/r & n/r \\
\bottomrule
\end{tabular}
\par\smallskip
{\footnotesize $^a$ Including the descriptive $\ell=8$ run. $^b$ Seeds 1 and 2 ran in one job step on
the same GPU; 4.88\,h is that step's time. $^c$ While scoring. The job mean was 4--15\%, because
the reproduction pipeline leaves the GPU idle while it prepares data on the CPU (2--4\,min model
load, 9--24\,min feature pooling per group).}
\end{table}

\paragraph{Code and data availability.} The following will be released publicly with the final version of this paper: the code (item builder, feature extraction, the socket pipeline with its fit,
scoring and analysis commands, the trained baselines, and the figure scripts); the frozen analysis
plan with its amendments; the \geneblind{} train and development manifests; the split manifest; and
the per-item development and test scores behind every number in this paper. The frozen plugs and per-host
text slices are released with SHA-256 sums. \geneblind{} is built only from open resources (Reactome,
Gene Ontology, UniProt, ClinVar, MANE). All hosts and encoders are public checkpoints at the
revisions listed in \cref{app:layer}.

\clearpage
\section{Full test-set table}
\label{app:test-full}

\begin{table}[h]
\centering
\footnotesize
\setlength{\tabcolsep}{3.5pt}
\caption{\textbf{Test-set results, all rows} (single run; 7,208 questions; 3,468 about genes with a related training
gene, 3,740 without). Gaps (\%) with 95\% confidence intervals. Controls are given against the empty write
(sham): within $\pm2\%$ means no effect. Modular plugs: the range over the three other models' plugs carried into
this model. Trained MLP: mean over training runs (interval in \cref{tab:test}).}
\label{tab:test-full}
\begin{tabular}{@{}lcccc@{}}
\toprule
 & \qfour{} $\ell6$ & \qtwentyseven{} $\ell17$ & \mistral{} $\ell5$ & \olmo{} $\ell3$ \\
\midrule
Accuracy: real / shuffled       & 41.8 / 19.6 & 46.0 / 19.9 & 36.8 / 20.1 & 38.7 / 20.2 \\
Gap [95\% CI]              & +22.2 \ci{19.8}{25.0} & +26.1 \ci{24.3}{27.9} & +16.7 \ci{14.2}{19.7} & +18.4 \ci{16.2}{21.0} \\
Significant; above 4\%            & reject, above & reject, above & reject, above & reject, above \\
Gap without \fdis{} questions                 & +23.5 \ci{20.8}{26.5} & +27.5 \ci{25.7}{29.3} & +17.7 \ci{15.2}{20.9} & +19.6 \ci{17.2}{22.3} \\
Gap, \fdis{} questions only               & +7.0 \ci{2.8}{11.4} & +9.0 \ci{4.4}{14.0} & +4.3 \ci{1.4}{7.2} & +4.9 \ci{1.5}{8.2} \\
Empty-write accuracy; shuffled $-$ empty & 20.0; $-0.4$ & 19.6; $+0.3$ & 19.8; $+0.3$ & 19.6; $+0.6$ \\
Re-paired ($\times5$); random map & $-0.8$ to $0.0$; $-0.5$ & $-1.0$ to $0.0$; $-0.5$ & $-1.1$ to $+0.3$; $-1.1$ & $-0.8$ to $+0.9$; $+0.1$ \\
Real $-$ another gene's DNA              & +21.9 \ci{19.4}{24.5} & +26.1 \ci{23.7}{28.8} & +16.9 \ci{14.4}{20.2} & +19.6 \ci{17.4}{22.2} \\
Plug $-$ nearest training gene     & +3.8 \ci{2.3}{5.2} & +4.1 \ci{1.6}{6.3} & +3.0 \ci{1.4}{4.5} & +5.2 \ci{3.8}{6.5} \\
Gap, genes with / without relatives       & +23.9 / +20.6 & +26.1 / +26.1 & +18.7 / +14.9 & +19.7 / +17.2 \\
Gap, refitted without relatives         & +19.3 \ci{17.8}{20.9} & +23.2 \ci{21.8}{24.6} & +14.5 \ci{12.6}{16.6} & +16.1 \ci{14.5}{17.8} \\
Nearest gene, without relatives    & +13.6 & +16.6 & +9.8 & +10.1 \\
Other models' plugs carried in         & +21.9 to +22.4 & +25.2 to +26.4 & +16.4 to +17.2 & +18.4 to +18.6 \\
Own text written, vs empty (ceiling) & +25.4 & +32.8 & +20.3 & +20.1 \\
Neighbouring block          & $\ell5$: +19.4 \ci{16.9}{22.3} & $\ell13$: +26.6 \ci{24.6}{28.6} & $\ell8$: +18.8 \ci{16.6}{21.4} & --- \\
Trained MLP adaptor (mean)              & +17.3 & +13.9 & --- & --- \\
\bottomrule
\end{tabular}
\end{table}

\section{Development (\devconf{}) results}
\label{app:devb}
\label{sec:res-socket}

Before the test, the socket was confirmed on \devconf{} at the frozen layers (\cref{tab:main}). It gave
$+23.8$ \ci{+20.2}{+27.3} on \qfour{} and $+26.0$ \ci{+22.4}{+29.3} on \qtwentyseven{} (both reject
after Holm, $p<0.0005$, above SESOI; \devtune{} agrees within 2\%). On \mistral{} it gave $+17.2$
\ci{+13.3}{+21.4}, and on \olmo{}, at its out-of-sample layer, $+20.7$ \ci{+17.4}{+23.8}. The few
\devconf{} controls outside the $\pm2$ band lay \emph{below} sham (\cref{app:controls}).

\begin{table}[h]
\centering
\small
\setlength{\tabcolsep}{5pt}
\caption{\textbf{Results on the development confirmation set} (1,464 questions), scored before the test under the
same plan. Read as \cref{tab:test}. The correction for several models covers the three models scored under the plan
here; \olmo{} entered it only on the test set. n/c: not computed.}
\label{tab:main}
\begin{tabular}{@{}lcccc@{}}
\toprule
 & \qfour{} & \qtwentyseven{} & \mistral{} & \olmo{} \\
\midrule
Block written after / blocks          & 6 / 32 & 17 / 64 & 5 / 32 & 3 / 32 \\
How the block was chosen                    & peak-margin & peak-margin & half-max (in sample) & half-max (held out) \\
Accuracy: empty / real / shuffled   & 19.3 / 43.6 / 19.7 & 22.7 / 47.8 / 21.9 & 20.0 / 38.7 / 21.4 & 21.7 / 40.3 / 19.6 \\
\textbf{Gap (real $-$ shuffled)}                  & \textbf{+23.8} \ci{20.2}{27.3} & \textbf{+26.0} \ci{22.4}{29.3} & \textbf{+17.2} \ci{13.3}{21.4} & \textbf{+20.7} \ci{17.4}{23.8} \\
Significant; above 4\%                 & reject, above & reject, above & reject, above & ---, above \\
Gap without \fdis{} questions                    & +25.2 \ci{21.7}{28.8} & +27.6 \ci{23.8}{31.2} & +18.3 \ci{14.1}{22.4} & +21.2 \ci{17.5}{24.7} \\
Real $-$ another gene's DNA                 & +24.5 \ci{20.9}{28.1} & +26.8 \ci{22.6}{30.8} & +17.6 \ci{13.4}{21.9} & +21.2 \ci{17.4}{24.8} \\
Plug $-$ nearest training gene        & +4.5 \ci{1.7}{7.3}    & +1.7 \ci{-2.0}{5.5}   & +3.6 \ci{0.6}{6.4}    & +6.2 \ci{3.2}{9.0} \\
Gap, refitted without relatives            & +20.4 \ci{17.0}{23.7} & +23.6 \ci{20.6}{26.6} & +15.7 \ci{11.9}{20.0} & +18.9 \ci{15.0}{22.6} \\
Gap, aimed at symbol / description    & +13.7 / +11.1 & +19.7 / +20.9 & +11.0 / +11.3 & +9.6 / +14.1 \\
Trained MLP adaptor (mean)        & +19.1 \ci{3.8}{34.4} & +14.5 \ci{3.3}{25.8} & --- & --- \\
\bottomrule
\end{tabular}
\end{table}